\documentclass[english]{svmult}
\usepackage[T1]{fontenc}
\usepackage[latin9]{inputenc}
\usepackage{textcomp}
\usepackage{url}
\usepackage{amsmath}
\usepackage{amssymb}
\usepackage{graphicx}
\usepackage{esint}

\makeatletter

\newcommand{\lyxmathsym}[1]{\ifmmode\begingroup\def\b@ld{bold}
 \text{\ifx\math@version\b@ld\bfseries\fi#1}\endgroup\else#1\fi}

\newcommand{\lyxdot}{.}

\usepackage{multicol}
\usepackage{cite}
\usepackage[bottom]{footmisc}

\usepackage{babel}
\begin{document}

\title{Theory of deflagration and fronts of tunneling in molecular magnets}

\author{D. A. Garanin}

\institute{Department of Physics and Astronomy, Lehman College, City University
of New York, 250 Bedford Park Boulevard West, Bronx, New York 10468-1589,
USA\texttt{ dmitry.garanin@lehman.cuny.edu}}

\maketitle

\abstract{Decay of metastable states in molecular magnets (MM) leads to energy
release that results in temperature increase that, in turn, positively
affects the decay rate. This is the mechanism of recently discovered
magnetic deflagration that is similar to regular chemical burning
and can propagate in a form of burning fronts in long MM crystals.
Near spin-tunneling resonances the decay rate is also affected by
the dipolar field (self-consistent with the switching magnetization)
that can block or unblock tunneling. There are non-thermal fronts
of tunneling in which the magnetization adjusts in such a way that
the system is on resonance within the front core, so that the tunneling
front can propagate. In general, both dipolar field and temperature
control fronts of quantum deflagration. The front speed can reach
sonic values if a strong transverse field is applied to boost tunneling.}

\section{Introduction}

\label{sec:Introduction}


Deflagration or burning is decay of metastable states accelerated
by the temperature rise due to the energy release in this process\cite{gla96book,lanlif9fluid}.
In most cases the decay rate has the Arrhenius temperature dependence,
$\Gamma=\Gamma_{0}\exp\left[-U/\left(k_{B}T\right)\right]$, where
$U$ is the energy barrier. Because of the very strong positive feedback,
burning can have a form of a thermal runaway: almost undetectably
slow relaxation at the beginning followed by an explosion-like relaxation
at the end (explosions at ammunition-storage sites, Bhopal disaster,
etc.). In other cases there is a burning front propagating with a
constant speed away from the ignition point. These fronts are driven
by the heat conduction from the hot burned region to the cold unburned
region before the front. Burning of a sheet of paper is a good example
of a deflagration front.

Molecular magnets (MM), of which the most famous is Mn$_{12}$Ac \cite{lis80},
are burnable materials because of their bistability resulting from
a strong uniaxial anisotropy that creates an energy barrier \cite{sesgatcannov93nat}.
One can make magnetic state metastable by applying a magnetic field
along the anosotropy axis. Burning, of course, should lead to a much
faster relaxation than a regular relaxation at fixed low temperatures.
Indeed, in early experiments on relaxation of large specimens of MM
\cite{paupark95kluwer,fometal97prl,baretal99prb} an abrupt and nearly
total relaxation of the metastable magnetization has been detected
but not explained. The 2005 space-resolved experiments of the Sarachik
group \cite{suzetal05prl} on long crystals of Mn$_{12}$Ac have shown
propagating fronts of relaxation. In this experiments, regularly-spaced
Hall probes at the sides of the crystal detected the transverse magnetic
field created by the non-uniformity of the magnetization \cite{avretal05prb}.
Chudnovsky interpreted these propagating fronts of relaxation as fronts
of deflagration \cite{suzetal05prl}. Measurements of the time dependence
of the total magnetization by the Tejada group, inspired by the above
experiment, have shown a linear time dependence that was attributed
to a deflagration front travelling through a Mn$_{12}$ crystal \cite{heretal05prl}.
Here quantum maxima of the front speed vs the bias field have been
detected, Fig. 4 of Ref. \cite{heretal05prl}. Discovery of magnetic
deflagration opened an active field of experimental research, mainly
on Mn$_{12}$Ac \cite{mchughetal07prb,hughetal09prb-species,hughetal09prb-tuning,mchughetal09prb}.
Experiments at high sweep rates \cite{vanetal04prbrc,decvanmostejhermac09prl}
have shown spin avalanches propagating at a fast speed. In this region,
deflagration can go over into detonation \cite{modbycmar11prl}. Magnetic
deflagration (coupled to a structural phase transition) has also been
observed on manganites \cite{masiaetal07prb} and intermetallic compounds
\cite{velezetal10prb,velezetal12prb}. To the contrast, it is problematic
to observe deflagration fronts on another popular MM Fe$_{8}$ because
of the pyramidal shape of its crystals.

One can ask if deflagration can exist in traditional magnetic systems,
many having a strong uniaxial anisotropy. Unfortunately, the energy
release in magnetic systems is much weaker than in the case of a regular
(chemical) deflagration. Thus, at room temperatures, the ensuing temperature
increase is too small to change the relaxation rate and support burning.
Only at low temperatures the increase of the relaxation rate becomes
large. A hallmark of magnetic deflagration is its non-destructive
character. ``Burned'' MM can be recycled (put again into the metastable
state) by simply reversing the longitudinal magnetic field.

A comprehensive theory of magnetic deflagration given in Ref. \cite{garchu07prb}
includes calculations of the stationary speed of the burning front,
ignition time due to local increase of temperature or change of the
magnetic field, as well as the analysis of stability of the low-temperature
state with respect to deflagration that depends on the heat contact
of the MM crystal with the environment. However, up to now there is
no complete accordance between the theory and experiment for several
reasons. First, thermal diffusivity $\kappa$ of Mn$_{12}$ that plays
a crucial role in deflagration has not been measured up to now. Second,
there is no completely satisfactory theory of relaxation in molecular
magnets that takes into account important collective effects such
as superradiance and phonon bottleneck.

Because of their not too large spin ($S=10$ for Mn$_{12}$ and Fe$_{8}$),
molecular magnets are famous exponents of spin tunneling \cite{chu79jetp,enzsch86,chugun88prl,chugun88prb}
that has a resonance character and leads to the steps in dynamic hysteresis
curves at the values of the longitudinal magnetic field where quantum
levels of the spin at the two sides of the potential barrier match
\cite{frisartejzio96prl,heretal96epl,thoetal96nat}. Since the discovery
of magnetic deflagration there was a quest for quantum effects in
it. The simplest approach \cite{heretal05prl,garchu07prb} uses the
fact that usually spin tunneling occurs via pairs of quantum levels
just below the classical barrier. This tunneling is thermally assisted
and can be described by an effective lowering of the energy barrier
at resonance values of the bias field (Fig. 2 of Ref. \cite{baretal99prb}).
Thus using the Arrhenius relaxation rate with such an effective barrier
does incorporate spin tunneling. Experimentally it was found that
spin tunneling strongly affects ignition of deflagration (Fig. 5 of
Ref. \cite{mchughetal07prb}) and to a smaller extent the front speed
(Fig. 5 of Ref. \cite{mchughetal07prb} and Fig. 4 of Ref. \cite{heretal05prl}).

Quantum effects in deflagration shoud be sensitive to the dipolar
field created by the sample. In a long uniformly magnetized crystal
of Mn$_{12}$Ac the dipolar field is $B^{(D)}=52.6$ mT, as calculated
microscopically in Ref. \cite{garchu08prb}, while the measured value
\cite{mchughetal09prb} is very close to it. This creates a dipolar
energy bias $W^{(D)}=g\mu_{B}B^{(D)}(m'-m)$ between the pair of resonant
quantum levels $m$ and $m'$ (quantum numbers for $S_{z}$ in the
two energy wells). This energy bias typically largely exceeds the
tunnel splitting $\Delta$ that contributes to the resonance width.
In the deflagration front the dipolar field typically changes between
$+B^{(D)}$ and $-B^{(D)}$ and so does the energy bias. As the result,
spin tunneling in the deflagration front does not occur at a fixed
resonance condition. This can explain why the observed quantum maxima
in the front speed can be not as strong as expected, compared to the
effect of tunneling on the ignition of deflagration.

Further theoretical research led to the idea of the dipole-dipole
interaction (DDI) playing an active role in deflagration by controlling
the relaxation rate, as temperature does in the regular deflagration.
Adding to the external bias field, the dipolar field can set particular
magnetic molecules on or off resonance, facilitating or blocking their
tunneling relaxation. The problem is self-consistent since tunneling
of one magnetic molecule changes dipolar fields on the other ones.
A numerical solution of this problem in a form of a moving front of
tunneling at zero temperature (sometimes called ``cold deflagration'')
has been found in Ref. \cite{garchu09prl}. An analytical solution
for the front of tunneling in the realistic strong-DDI case has been
obtained in Ref. \cite{gar09prb}.

Pure non-thermal fronts of tunneling can occur in the case of a very
good thermal contact of the MM crystal with the environment, so that
its temperature does not increase and remains so low that tunneling
takes place directly from the metastable ground state into a matching
excited state on the other side of the barrier. This process can be
efficient only if a strong transverse field is applied and the corresponding
tunnel splitting $\Delta$ is large enough. In this case the speed
of fronts of tunneling can theoretically exceed the speed of a regular
deflagration by a large margin. Indeed, the dipolar field in the crystal
changes instantaneously, in contrast to the temperature changing via
heat conduction. Second, the relaxation rate due to tunneling directly
from the ground state can be much higher than the relaxation rate
due to the barrier-climbing processes in the regular deflagration.

If the MM crystal is thermally insulated, its temperature is increasing
as a result of a decay of the metastable state, so that there can
be a mixture of both mechanisms of deflagration considered above \cite{garjaa10prbrc}.
Whereas far from resonances a regular deflagration takes place, near
resonances tunneling leads to a great increase of the front speed.
A more detailed treatment of the quantum-thermal deflagration for
a realistic model of Mn$_{12}$Ac with $S_{z}^{4}$ terms in the effective
Hamiltonian is given in recent Ref. \cite{garsho12prb}.

Theories of fronts of tunneling mentioned above are based on the model
simplification considering it as one dimensional. In the regular deflagration,
there is a mechanism that makes fronts flat and smooth (laminar),
so that the deflagration problem in long crystals indeed becomes 1$d$.
In the case of dipolar-driven fronts of tunneling, it is not immediately
clear whether fronts are flat or not, and, moreover, there is a mechanism
that favors non-laminar fronts. The full 3$d$ theory of fronts of
tunneling that will be presented below, numerically yields non-flat
and non-laminar fronts. The latter slows down the front speed in comparison
to the simplified 1$d$ theory but, nevertheless, the speed can reach
values comparable with the speed of sound in MM near tunneling resonances
in strong transverse field.

In the main part of this contribution, first the regular (thermal)
magnetic deflagration will be considered. Then calculation of the
dipolar field in molecular magnets will be explained. The final part
is devoted to the theory of fronts of tunneling.

\section{Magnetic deflagration\label{sec:Deflagration}}

\begin{figure}[t]
\centerline{\includegraphics[angle=0,width=8cm]{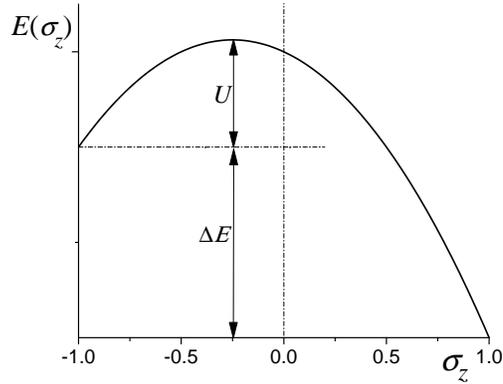}}
\caption{Energy barrier of a biased molecular magnet, $\sigma_{z}\equiv S_{z}/S$.}
\label{fig:Barrier}
\end{figure}

For the generic model of a molecular magnet the energy has the form
\begin{equation}
\mathcal{H}=-DS_{z}^{2}-g\mu_{B}B_{z}S_{z}+\mathcal{H}^{\prime},\label{eq:Ham}
\end{equation}
where $D>0$ is the uniaxial anisotropy constant and $\mathcal{H}^{\prime}$
stands for all terms that do not commute with $S_{z}$ and thus cause
spin tunneling. In Mn$_{12}$Ac there is an additional smaller longitudinal
term $-AS_{z}^{4}$, the implications of which will be discussed later.
In the biased case $B_{z}>0$, the dependence of the energy on $\sigma_{z}\equiv S_{z}/S$
is sketched in Fig. \ref{fig:Barrier}. The energy barrier $U$ shown
in Fig. \ref{fig:Barrier} has the form
\begin{equation}
U=\left(1-h\right)^{2}U_{0},\qquad U_{0}=DS^{2},\qquad h\equiv g\mu_{B}B_{z}/(2DS).\label{eq:UDef}
\end{equation}
With $S=10$ the zero-field energy barrier $U_{0}$ has a large value
of 67 K in Mn$_{12}$Ac. The energy of the metastable state is given
by $\Delta E=2Sg\mu_{B}B_{z}$.

In the absence of spin tunneling at low temperatures, $U/\left(k_{B}T\right)\gg1$,
the rate equation describing relaxation of the metastable population
$n$ (the fraction of magnetic molecules in the left well) has the
form
\begin{equation}
\dot{n}=-\Gamma\left(n-n^{(\mathrm{eq})}\right),\label{eq:ndotEq}
\end{equation}
where the relaxation rate is given by
\begin{equation}
\Gamma=\Gamma_{0}\exp\left(-\frac{U}{k_{B}T}\right)\left[1+\exp\left(-\frac{\Delta E}{k_{B}T}\right)\right].\label{eq:GammaDef}
\end{equation}
Here the second term in the square brackets describes back transitions
from the stable well to the metastable well. In the strong-bias case,
$\Delta E\gg k_{B}T$, this term can be omitted. The equilibrium metastable
population $n^{(\mathrm{eq})}$ is given by
\begin{equation}
n_{-}^{(\mathrm{eq})}=1/\left[\exp\left(\frac{\Delta E}{k_{B}T}\right)+1\right].\label{eq:neqDef}
\end{equation}
In the strong-bias case it can be neglected.

The second equation describing deflagration is the heat conduction
equation
\begin{equation}
C\dot{T}=\nabla\cdot k\nabla T-\dot{n}\Delta E,\label{eq:TdotEq}
\end{equation}
where $k$ is thermal conductivity and $C$ is heat capacity. The
second term on the right is the energy release due to decay of the
metastable state. The heat capacity is mainly due to phonons, whereas
the magnetic contribution is relatively small. At low temperatures
only acoustic phonons are excited, whereas high-energy optical phonons
are frozen out, thus $C$ has the form \cite{kit63}
\begin{equation}
C=Ak_{B}\left(T/\Theta_{D}\right)^{\alpha},\label{eq:Cph}
\end{equation}
where $\alpha=3$ in three dimensions, $A=12\pi^{4}/5\simeq234$ is
a numerical factor and $\Theta_{D}$ is the Debye temperature, $\Theta_{D}\simeq40$
K for Mn$_{12}$Ac. Although at low temperatures this expression is
in a reasonable accordance with measurements on Mn$_{12}$Ac \cite{gometal98prb},
its applicability range is very narrow, $T\lesssim5$ K. On the other
hand, the temperature generated in the deflagration (the so-called
flame temperature) is typically above 10 K. The heat capacity of Mn$_{12}$Ac
can be well described within a broad temperature range with the help
of the extended Debye model (EDM) \cite{gar08prbrc} that comprises
three different acoustic phonon modes as well as optical modes. Practically,
one can use measured values of $C$ \cite{gometal98prb}.

It is convenient to use the relation $C=d\mathcal{E}/dT$ to rewrite
Eq. (\ref{eq:TdotEq}) in terms of the energy $\mathcal{E}$ as
\begin{equation}
\dot{\mathcal{E}}=\nabla\cdot\kappa\nabla\mathcal{E}-\dot{n}\Delta E,\label{eq:EdotEq}
\end{equation}
where $\kappa=k/C$ is thermal diffusivity. The latter has not yet
been measured, although a crude estimate $\kappa\simeq10^{-5}$ m$^{2}$/s
was deduced from experiments \cite{suzetal05prl,hughetal09prb-tuning}.
This value is comparable with that of metals. Temperature dependence
of $\kappa$ that could be substantial at low temperatures remains
unknown.

Equations (\ref{eq:ndotEq}) and (\ref{eq:EdotEq}), together with
Eq. (\ref{eq:GammaDef}) and the relation
\begin{equation}
\mathcal{E}(T)=\int_{0}^{T}C(T')dT',\label{eq:EviaT}
\end{equation}
is a strongly-nonlinear system of equations. It is easy to solve these
equations numerically but it costs efforts to do it analytically.
The two main problems to solve are (i) stability of the low-temperature
state with respect to thermal runaway or ignition of a deflagration
front and (ii) the shape and speed of the stationary deflagration
front in long crystal.

\subsection{Ignition of deflagration\label{sub:Ignition}}

\begin{figure}[t]

\centering\includegraphics[angle=0,width=8cm]{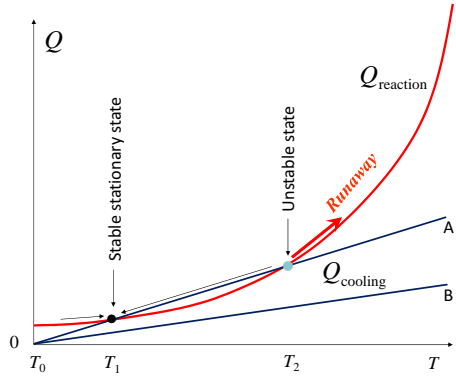}\caption{Semenov's mechanism of a thermal runaway, Eq. (\ref{eq:Semenov}).}
\label{Fig:Semenov}
\end{figure}

If the sample is perfectly thermally insulated, the whole released
energy remains inside and the temperature monotonically increases.
This leads to a thermal instability that can take a considerable time
to develop, the ignition time. If there is a thermal contact with
the environment, maintained at a constant low temperature $T_{0}$,
there are two possible cases. In the subcritical case, the temperature
rise in the sample due to slow decay leads to temperature gradients
and heat flow toward the sample boundaries that ensures a stationary
low-temperature state (proper conditions of explosives' storage).
In the supercritical case, heat loss through the boundaries is insufficient
to balance the increase of the heat release due to rise of temperature.
This leads to ignition of a self-supporting burning process. In small
crystals of MM, temperature gradients are higher and heat loss to
the environment is more efficient. In larger crystals, temperature
gradients are lower and thermal instability is more likely. This is
why deflagration was observed in larger crystals.

Thermal instability occurs because of a stronger temperature dependence
of the relaxation rate, Eq. (\ref{eq:GammaDef}), than that of the
heat exchange with the environment. The essence of the problem is
contained in the old model of explosive instability by Semenov described
by a single equation
\begin{equation}
\dot{T}=Q_{\mathrm{reaction}}-Q_{\mathrm{cooling}},\label{eq:Semenov}
\end{equation}
where $Q_{\mathrm{reaction}}\sim\Gamma(T)$ and $Q_{\mathrm{cooling}}=\alpha(T-T_{0})$.
In the case B in Fig. \ref{Fig:Semenov}, the thermal contact to the
bath is too weak, $Q_{\mathrm{cooling}}<Q_{\mathrm{reaction}}$ at
all $T$, so that the system is absolutely unstable. In the case A,
the thermal contact is stronger and there is a stability range $T<T_{2}$,
where the stationary state $T=T_{1}$ is an attractor. However, heating
the system above $T_{2}$ leads to thermal explosion.

Semenov's model is zero-dimensional, whereas in MM crystals the problem
is at least one-dimensional and more complicated. There are different
cases of thermal instability, mainly instability of a large crystal
initially at uniform temperature (that begins at the center), instability
due to heating one end of a long crystal, and the instability due
a magnetic field gradient that makes the barrier lower at one side
of the crystal. Analysis of all these cases has been done in Ref.
\cite{garchu07prb}. In particular, when the magnetic field and temperature
of the sample boundary $T_{0}$ are independent of coordinates, the
crystal loses stability against formation and propagation of the flame
(magnetic avalanche) when the rate of the spin flip for an individual
molecule, $\Gamma(H,T_{0})$ , exceeds
\begin{equation}
\Gamma_{c}=\frac{k_{B}T_{0}}{U}\frac{8kT_{0}}{l^{2}n_{i}\Delta E}\,.\label{eq:GammacIgnition}
\end{equation}
Here $k$ is thermal conductivity at $T_{0}$ and the length parameter
$l$ is uniquely determined by geometry, being of the order of the
smallest dimension of the crystal, whereas $n_{i}$ is the metastable
population in the initial state.

Experimentally magnetic deflagration can be initiated either by heating
one end of the crystal \cite{mchughetal07prb,hughetal09prb-species,hughetal09prb-tuning}
or by sweeping the magnetic field in the positive direction, that
reduces the energy barrier and makes the condition in Eq. (\ref{eq:GammacIgnition})
satisfied \cite{suzetal05prl}. In Ref. \cite{heretal05prl} deflagration
was ignited by surface acoustic waves (SAW), instead of heating.

\subsection{Deflagration fronts\label{sub:Deflagration-fronts}}

Fronts of magnetic burning propagating in long crystals of molecular
magnets are flat and smooth, i.e., the problem of deflagration is
one-dimensional. The stability of flat fronts can be immediately seen.
Indeed, if a fraction of a front gets ahead of neighboring fractions,
the heat released at this place will be propagating not exactly straight
ahead (as in a flat front) but also sideways. This will slow down
this leading fraction of the front and speed up the lagging fractions
surrounding it. Thus any local deviation from a flat front will disappear
with time.

In a stationary-moving front, all physical quantities depend only
on the combined variable that can be chosen, e.g., in the time-like
form $u\equiv t-z/v$, where $v$ is the front speed. In terms of
$u$ the deflagration equations have the form
\begin{eqnarray}
\frac{dn}{du} & = & -\Gamma(T)\left(n-n^{(\mathrm{eq})}(T)\right)\nonumber \\
\frac{d\mathcal{E}}{du} & = & \frac{1}{v^{2}}\frac{d}{du}\kappa\frac{d\mathcal{E}}{du}-\frac{dn}{du}\Delta E\label{eq:nEdotEqsu}
\end{eqnarray}
plus Eq. (\ref{eq:EviaT}). Integrating the energy equation one obtains
\begin{equation}
\mathcal{E}+n\Delta E-\frac{\kappa}{v^{2}}\frac{d\mathcal{E}}{du}=\mathrm{const}.\label{eq:EnIntegral}
\end{equation}
Far before and far behind the front, the term with the derivative
vanishes. Thus one obtains the energy conservation law in the form
\begin{equation}
\mathcal{E}_{i}+n_{i}\Delta E=\mathcal{E}_{f}+n^{(\mathrm{eq})}(T_{f})\Delta E,\label{eq:EnergyConservation}
\end{equation}
where $i$ stands for ``initial'' (before the front) and $f$ stands
for ``final'' or ``flame''. This is a transcedental equation for
the flame temperature $T_{f}$ that has to be solved together with
Eq. (\ref{eq:EviaT}). If $\mathcal{E}_{i}\approx0$ (low initial
temperature) and $n^{(\mathrm{eq})}(T_{f})$ is negligible (full-burning
case realised at a strong bias, see Eq. (74) of Ref. \cite{garchu07prb})
one immediately finds the flame energy from $n_{i}\Delta E=\mathcal{E}_{f}$,
and then $T_{f}$ follows by inverting Eq. (\ref{eq:EviaT}). In the
incomplete-burning regime at small bias, a pulsating instability of
stationary deflagration fronts \cite{modbycmar11prb} was found. The
operations above assume that the heat is not exchanged via the sides
of the crystal. In the opposite case, the energy conservation becomes
invalid and the theory has to be extended.

One can immediately get an idea of the front speed by rewriting the
deflagration equations (\ref{eq:nEdotEqsu}) in the dimensionless
form. In terms of the reduced variables
\begin{equation}
\tilde{n}\equiv n/n_{i},\qquad\tilde{\mathcal{E}}\equiv\mathcal{E}/(n_{i}\Delta E),\qquad\tilde{u}\equiv u\Gamma_{f}\label{eq:ReducedVars}
\end{equation}
and parameters
\begin{equation}
\tilde{\Gamma}\equiv\Gamma/\Gamma_{f},\qquad\tilde{\kappa}\equiv\kappa/\kappa_{f}\label{eq:ReducedPars}
\end{equation}
equations (\ref{eq:nEdotEqsu}) become
\begin{eqnarray}
\frac{d\tilde{n}}{d\tilde{u}} & = & -\Gamma\left(\tilde{n}-\tilde{n}^{(\mathrm{eq})}\right)\nonumber \\
\frac{d\tilde{\mathcal{E}}}{d\tilde{u}} & = & \frac{1}{\tilde{v}^{2}}\frac{d}{d\tilde{u}}\tilde{\kappa}\frac{d\tilde{\mathcal{E}}}{d\tilde{u}}-\frac{d\tilde{n}}{d\tilde{u}},\label{eq:ReducedEqs}
\end{eqnarray}
where the reduced front speed $\tilde{v}$ is related to the actual
front speed $v$ by
\begin{equation}
v=\tilde{v}\sqrt{\kappa_{f}\Gamma_{f}}.\label{eq:v_via_vtil}
\end{equation}
Refs. \cite{lanlif9fluid,suzetal05prl} give the expression above
without $\tilde{v}$ for the front speed.

It turns out that $\tilde{v}$ in Eq. (\ref{eq:v_via_vtil}) is not
merely a number but rather it is a function of dimensionless parameters
such as
\begin{equation}
W_{f}\equiv U/(k_{B}T_{f}).\label{eq:WfDef}
\end{equation}
 Because of the non-linearity of Eq. (\ref{eq:ReducedEqs}), their
general analytical solution that defines $\tilde{v}$ does not exist.
There are two parameter ranges in the problem: Slow-burning high-barrier
range $W_{f}\gg1$ and fast-burning low-barrier range $W_{f}\lesssim1$.

In the former, burning occurs in the front region where the temperature
is already close to $T_{f}$. Assuming that $\kappa$ is temperature
independent, $\tilde{\kappa}=1$, and linearizing $\Gamma(T)$ near
$T_{f}$, one can solve the problem analytically. Within the full-burning
approximation ($n^{(\mathrm{eq})}\Rightarrow0$) the reduced front
speed is given by \cite{garchu07prb}
\begin{equation}
\tilde{v}=\sqrt{\frac{C_{f}T_{f}}{n_{i}\Delta E}\frac{k_{B}T_{f}}{U}}.\label{eq:vtilRes}
\end{equation}
With the help of Eq. (\ref{eq:Cph}) (that is not accurate, however!)
this result simplifies to
\begin{equation}
\tilde{v}=\sqrt{(\alpha+1)/W_{f}}.\label{eq:vtilalpha}
\end{equation}
The applicability range of these expressions is $\tilde{v}\ll1$.

The corresponding profile of the metastable population $n$ in the
front has the form
\begin{equation}
\tilde{n}=\frac{1}{1+e^{u}}=\frac{1}{2}\left(1-\tanh\frac{\tilde{u}}{2}\right)\label{eq:ntilviautilfront}
\end{equation}
that corresponds to the symmetric $\tanh$ magnetization profile $\sigma_{z}=1-2n=\tanh(\tilde{u}/2)$.
In real units the result reads
\begin{equation}
n=\frac{n_{i}}{2}\left[1+\tanh\left(\frac{z}{2\tilde{v}l_{d}}-\frac{\Gamma_{f}t}{2}\right)\right],\label{eq:nviaxtfront}
\end{equation}
where $l_{d}=\sqrt{\kappa_{f}/\Gamma_{f}}$ is the \textit{a-priori}
with of the deflagration front. Magnetization profile of this kind
can be seen in Fig. 11 of Ref. \cite{garchu07prb} and in the upper
panel of Fig. 10 of Ref. \cite{garsho12prb}. The reduced energy in
the front is given by
\begin{equation}
\tilde{\mathcal{E}}=(1-e^{-u})^{-\tilde{v}^{2}}=(1-\tilde{n})^{\tilde{v}^{2}}.\label{eq:Etilfront}
\end{equation}
Since in the high-barrier approximation $\tilde{v}\ll1$, the formula
above yields $\tilde{\mathcal{E}}\approx1$ in the active burning
region and actually everywhere except for the region far ahead of
the front where $\tilde{n}$ is very close to 1. This justifies the
approximation made.

It should be noted that the full-burning approximation used above
requires a bias high enough thus the barrier low enough, $W_{f}\lesssim6$,
according to Eq. (79) of Ref. \cite{garchu07prb}. Thus the applicability
range of the slow-burning high-barrier approximation is rather limited.
The theory can be improved by taking into account incomplete burning.
However, this makes analytics cumbersome because of the transcedental
equation (\ref{eq:EnIntegral}) defining $T_{f}$. Numerical solution
for the deflagration front poses no problems, nevertheless. Because
of incomplete burning, $T_{f}$ and thus the front speed decrease
below the values given above.

In the low-barrier fast-burning regime $W_{f}\lesssim1$ there is
no rigorous analytical solution to the problem. Additionally, the
Arrhenius form of the relaxation rate, Eq. (\ref{eq:GammaDef}), becomes
invalid. In this regime the magnetization profile is asymmetric, as
can be seen in the upper panel of Fig. 12 of Ref. \cite{garsho12prb}.

\begin{figure}
\centering\includegraphics[angle=0,width=8cm]{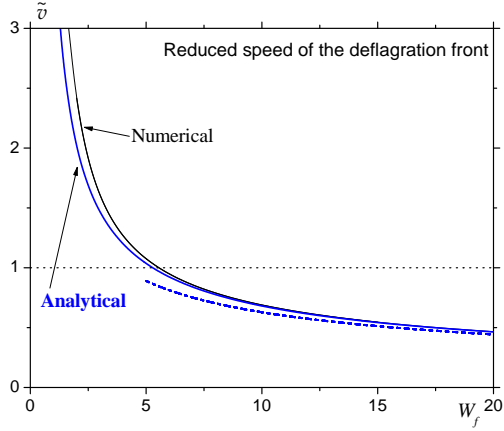}

\caption{Reduced speed of a deflagration front defined by Eq. (\ref{eq:v_via_vtil}).
The numerical result has been obtained in Ref. \cite{garchu07prb}
within the full-burning approximation using the low-temperature form
of the heat capacity, Eq. (\ref{eq:Cph}). Analytical result is Eq.
(\ref{eq:vtil_general_final}). The dotted line is the high-$W_{f}$
asymptote, Eq. (\ref{eq:vtilalpha}). }

\label{Fig:vtil-deflagration}
\end{figure}

Making the symplifying approximation for the relaxation rate
\begin{equation}
\tilde{\Gamma}(\tilde{\mathcal{E}})=\begin{cases}
0, & \tilde{\mathcal{E}}<\tilde{\mathcal{E}}_{0}\\
1, & \tilde{\mathcal{E}}>\tilde{\mathcal{E}}_{0},
\end{cases}\label{eq:GammaSimplified}
\end{equation}
where $\tilde{\mathcal{E}}_{0}$ will be defined below, one can solve
the problem of a stationary deflagration front in the whole parameter
range. Let us search for the front in which $\tilde{\mathcal{E}}=\tilde{\mathcal{E}}_{0}$
at $u=0$. In the reduced form of the energy equation (\ref{eq:EnIntegral}),
\begin{equation}
\frac{d\tilde{\mathcal{E}}}{d\tilde{u}}=\tilde{v}^{2}(\tilde{\mathcal{E}}+\tilde{n}-1),\label{eq:E_Eq_Integral_Reduced}
\end{equation}
one has $\tilde{n}=1$ before the front, $u<0$. Thus here the energy
equation solves to
\begin{equation}
\tilde{\mathcal{E}}=\tilde{\mathcal{E}}_{0}e^{\tilde{v}^{2}\tilde{u}}.
\end{equation}
On the other hand, for $u>0$ the solution of the population equation
$d\tilde{n}/d\tilde{u}=-\tilde{\Gamma}\tilde{n}=-\tilde{n}$ reads
$\tilde{n}=e^{-\tilde{u}}.$ Inserting this into Eq. (\ref{eq:E_Eq_Integral_Reduced}),
and integrating the differential equation, one obtains the solution
\begin{equation}
\tilde{\mathcal{E}}=\left(\tilde{\mathcal{E}}_{0}-\frac{1}{1+\tilde{v}^{2}}\right)e^{\tilde{v}^{2}\tilde{u}}+1-\frac{\tilde{v}^{2}}{1+\tilde{v}^{2}}e^{-\tilde{u}}.
\end{equation}
The first term of this expression must vanish because of the boundary
condition $\tilde{\mathcal{E}}(\infty)=1$. This defines the reduced
front speed,
\begin{equation}
\tilde{v}=\sqrt{\frac{1}{\tilde{\mathcal{E}}_{0}}-1}.\label{eq:vtil_via_Etil0}
\end{equation}

To define $\tilde{\mathcal{E}}_{0}$, consider the reduced Arrhenius
relaxation rate
\begin{equation}
\tilde{\Gamma}=\exp\left[W_{f}\left(1-\frac{1}{\tilde{T}}\right)\right]\label{eq:Gamma_Arrhenius_tilde}
\end{equation}
and require
\begin{equation}
W_{f}\left(1-\frac{1}{\tilde{T_{0}}}\right)=-1
\end{equation}
as the switching point between $\tilde{\Gamma}=0$ and $\tilde{\Gamma}=1$.
This yields
\begin{equation}
\tilde{T}_{0}=\frac{W_{f}}{1+W_{f}}.
\end{equation}
Using Eq. (\ref{eq:Cph}), one obtains
\begin{equation}
\tilde{\mathcal{E}}_{0}=\tilde{T}_{0}^{\alpha+1}=\left(\frac{W_{f}}{1+W_{f}}\right)^{\alpha+1}.
\end{equation}
Substituting this into Eq. (\ref{eq:vtil_via_Etil0}), one finally
obtains
\begin{equation}
\tilde{v}=\sqrt{\left(\frac{1+W_{f}}{W_{f}}\right)^{\alpha+1}-1}.\label{eq:vtil_general_final}
\end{equation}
Limiting cases of this formula are
\begin{equation}
\tilde{v}\cong\begin{cases}
\sqrt{(\alpha+1)/W_{f}}, & W_{f}\gg1\\
1/W_{f}^{(\alpha+1)/2}, & W_{f}\ll1.
\end{cases}\label{eq:vtil_general_limiting_cases}
\end{equation}
It is remarkable that the rigorously obtained high-barrier slow-burning
result of Eq. (\ref{eq:vtilalpha}) is recovered exactly. In the low-barrier
fast-burning case the reduced front speed becomes large, as well as
the actual front speed of Eq. (\ref{eq:v_via_vtil}). One can see
that Eq. (\ref{eq:vtil_general_final}) is in a good accordance with
the numerical solution shown in Fig. \ref{Fig:vtil-deflagration}.

The high-speed regime of the deflagration should be superceded by
detonation when the front speed approaches the speed of sound. In
detonation, thermal expansion resulting from burning sends a shock
wave into the cold region before the front. As a consequence, the
temperature before the front rises as a result of compression, initiating
burning. Such a mechanism was recently considered for Mn$_{12}$Ac
in Ref. \cite{modbycmar11prl}.

\section{Fronts of tunneling\label{sec:Fronts-of-tunneling}}

\subsection{Tunneling effects in the relaxation rate\label{sub:Relaxation_rate}}

The relaxation rate $\Gamma$ including spin tunneling is at the foundation
of the quantum theory of deflagration in molecular magnets. In the
generic model of MM, Eq. (\ref{eq:Ham}), tunneling resonances occur
at the values of the total bias field $B_{\mathrm{tot},z}$ (including
the self-produced dipolar field) equal to
\begin{equation}
B_{k}=kD/(g\mu_{B}),\qquad k=0,\pm1,\pm2,\ldots\label{eq:BzkDef}
\end{equation}
for all the resonances. Spin tunneling leads to the famous steps in
the dynamic hysteresis curves \cite{frisartejzio96prl,heretal96epl,thoetal96nat}.
In the real Mn$_{12}$Ac there is an additional term $-AS_{z}^{4}$
that makes higher-energy resonances be achieved at smaller $B_{z}$
than low-energy resonances. The resulting tunneling multiplets
\begin{equation}
g\mu_{B}B_{km}=k\left[D+\left(m^{2}+(m+k)^{2}\right)A\right]\label{eq:Bzkm-Def}
\end{equation}
were used to experimentally monitor \cite{bokkenwal00prl,wermugchr06prl}
the transition between thermally assisted and ground-state tunneling
\cite{chugar97prl} in Mn$_{12}$Ac. Below $B_{k}$ will stand for
the resonance field $B_{km}$, for simplicity of notations.

In the case of an isolated magnetic molecule, the probability of a
spin to be in one of the resonant quantum states is oscillating with
time with the frequency $\Delta/\hslash$, where $\Delta$ is the
tunnel splitting. However, coupling to the environment, e.g., to phonons,
introduces damping to these oscillatins. If the decay rate of at least
one of the resonance states, $\Gamma_{m}$ or $\Gamma_{m'}$, exceeds
$\Delta/\hslash$, tunneling oscillations of the spin are overdamped.
This can be illustrated in the case of a resonance between the metastable
ground state $\left|-S\right\rangle $ and the matching excited state
at the other side of the barrier $\left|m'\right\rangle $ of a biased
MM at zero temperature. Ignoring all other levels, that is justified
at $T=0$, one can write down the Schrödinger equation in the form
\cite{gar09prb}
\begin{eqnarray}
\dot{c}_{-S} & = & -\frac{i}{2}\frac{\Delta}{\hbar}c_{m^{\prime}}\nonumber \\
\dot{c}_{m^{\prime}} & = & \left(\frac{iW}{\hbar}-\frac{1}{2}\Gamma_{m^{\prime}}\right)c_{m^{\prime}}-\frac{i}{2}\frac{\Delta}{\hbar}c_{-S},\label{eq:Damped_Schroedinger}
\end{eqnarray}
where
\begin{equation}
W\equiv\varepsilon_{-S}-\varepsilon_{m^{\prime}}=(S+m')g\mu_{B}(B_{\mathrm{tot},z}-B_{k})\label{eq:W-bias-Def}
\end{equation}
is the energy bias between the two levels. Whereas the level $\left|-S\right\rangle $
is undamped, the level $\left|m'\right\rangle $ can decay into lower-lying
levels in the same well via phonon-emission processes. At $T=0$ there
are no incoming relaxation processes for $\left|m'\right\rangle $.
In this case the damped Schrödinger equation above is accurate, as
it can be shown to follow from the density matrix equation. In the
underdamped case $\Gamma_{m^{\prime}}\lesssim\Delta/\hbar$ the solution
of these equations is oscillating. The first choice for studying tunneling
dynamics in molecular magnets is the overdamped case $\Gamma_{m^{\prime}}\gg\Delta/\hbar$,
since for not too strong transverse fields $B_{\bot}$ the tunnel
splitting$\Delta$ is a high power of $B_{\bot}$ (Ref. \cite{gar91jpa})
and typically it is much smaller than $\Gamma_{m^{\prime}}$. In the
overdamped case the variable $c_{m^{\prime}}$ in Eq. (\ref{eq:Damped_Schroedinger})
adiabatically adjusts to the instantaneous value of $c_{-S}$ and
the solution greatly simplifies. Setting $\dot{c}_{m^{\prime}}=0$
in the second of these equations, one obtains
\begin{equation}
c_{m^{\prime}}=\frac{\Delta}{2\hbar}\frac{c_{-S}}{W/\hbar+i\Gamma_{m^{\prime}}/2}.\label{eq:c_mprime_eliminated}
\end{equation}
Inserting this into the first of equations (\ref{eq:Damped_Schroedinger})
yields a closed differential equation for $c_{-S}$. Using $n=\left|c_{-S}\right|^{2}$
for the metastable occupation number, one arrives at the rate equation
\begin{equation}
\dot{n}=-\Gamma n,\label{eq:n_rate_Eq}
\end{equation}
where the dissipative resonance-tunneling rate $\Gamma$ is given
by \cite{garchu97prb}
\begin{equation}
\Gamma=\frac{\Delta^{2}}{2\hbar^{2}}\frac{\Gamma_{m^{\prime}}/2}{\left(W/\hbar\right)^{2}+\left(\Gamma_{m^{\prime}}/2\right)^{2}}.\label{eq:Gamma_resonant}
\end{equation}
This is a Lorentzian function with the maximum at the resonance, $W=0$.
Eqs. (\ref{eq:n_rate_Eq}) and (\ref{eq:Gamma_resonant}) were used
in Refs. \cite{garchu09prl,gar09prb} to study dipolar-controlled
fronts of tunneling at $T=0$, or ``cold deflagration''. The full
system of equations (\ref{eq:Damped_Schroedinger}) could also be
used to this purpose but nothing had been published up to date.

At nonzero temperatures, tunneling transitions via higher energy level
pairs become possible (thermally-assisted tunneling) and one has to
take into account non-resonant thermal transitions over the top of
the barrier. This makes the problem more complicated, and one needs
to use the density matrix equation (DME) taking into account spin-phonon
interactions explicitly. One of the first works using DME for Mn$_{12}$Ac
was Ref. \cite{garchu97prb} in which spin tunneling was considered
with the help of the high-order perturbation theory \cite{gar91jpa}
for a small transverse field $B_{\bot}$. The spin-phonon processes
taken into account were due to dynamic tilting of the anisotropy axis
by transverse phonons. Ref. \cite{garchu97prb} could qualitatively
explain thermally-assisted tunneling via the level pairs just below
the classical barrier. However, tunneling via low-lying resonant level
pairs or tunneling directly out of the metastable ground state are
inaccessible by this method because large enough splitting requires
non-perturbatively large transverse field that can only be dealt with
numerically.

Further work on spin-phonon relaxation in MM lead to elucidation of
the universal relaxation mechanism \cite{chu04prl,chugarsch05prb}.
This mechanism consists in distortionless rotation of the crystal
field acting on a magnetic molecule, actually the same mechanism as
used in Ref. \cite{garchu97prb}. It was, however, understood that
this mechanism does not require any poorly-known spin-lattice coupling
constants and everything can be expressed through much easier accessible
crystal-field parameters. This mechanism was overlooked in older theories
of spin-lattice relaxation based on energy contributions responsible
for the coupling. Rotations, to the contrary, cost no energy and the
effect has a purely inertial origin.

\begin{figure}
\centering\includegraphics[angle=0,width=8cm]{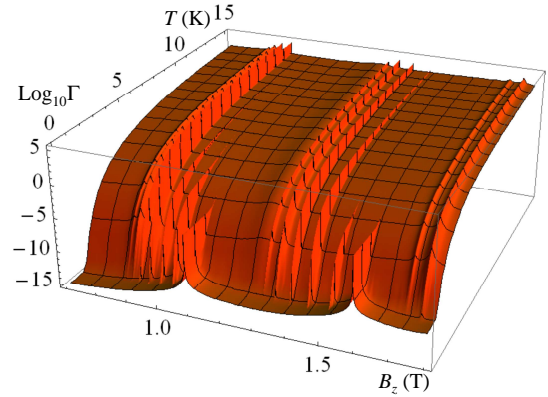}

\caption{Relaxation rate of Mn$_{12}$Ac vs temperature and longitudinal magnetic
field in a small transverse field. Resonance multiplets with $k=2,3$
are seen.}

\label{Fig:Gamma-3d-Javier}
\end{figure}

\begin{figure}
\centering\includegraphics[angle=-90,width=8cm]{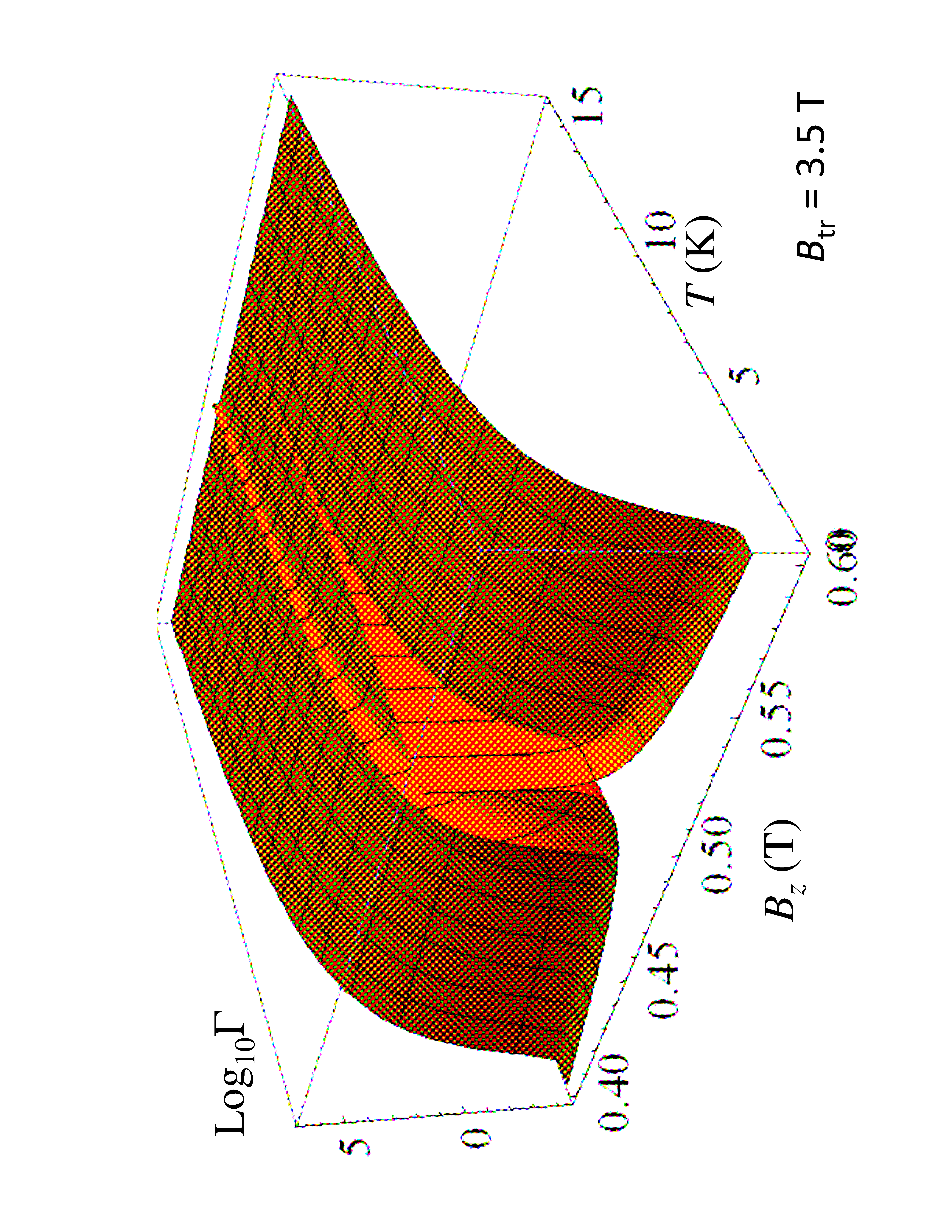}

\caption{Relaxation rate of Mn$_{12}$Ac vs temperature and longitudinal magnetic
field in the transverse field $B_{\bot}=3.5$ T. One can see the ground-state
resonance at $B_{z}=0.522$T and the first-excited-state resonance
at $B_{z}=0.490$T for $k=1$ multiplet.}

\label{Fig:Gamma-3d-Bz=00003D3.5}
\end{figure}

The universal relaxation mechanism allows a general numerical implementation
of the DME fully based on the crystal field parameters, recently accomplished
in Ref. \cite{gar12acp} that summarizes the current state of the
problem. Another important feature of Ref. \cite{gar12acp} is using
the so-called semi-secular approach capable of dealing with resonant
pairs of levels and thus describe spin tunneling. Conventional implementations
of the DME (see, e.g., Ref. \cite{luibarfer98prb}) use the secular
approach that crashes on tunneling resonances. In Ref. \cite{gar12acp}
the relaxation rate $\Gamma$ is extracted from the time-dependent
numerical solution of the DME (expressed in terms of eigenvalues and
eigenfunctions of the density matrix) as the inverse of the integral
relaxation time \cite{garishpan90tmf,gar96pre}. Unlike using the
lowest eigenvalue of the density matrix, this method also works at
elevated temperatures.

The temperature and field dependence of $\Gamma$ in Mn$_{12}$Ac
at a small transverse field ($B_{\bot}=0.04$T that typically arises
due to a 1º misalignment of the easy axis and the applied longitudinal
field) is shown in Fig. \ref{Fig:Gamma-3d-Javier}. One can see very
narrow and high maxima of $\Gamma$ (note that $\log\Gamma$ is plotted!)
due to spin tunneling. Maxima corresponding to the ground-state tunneling,
for which the maximum in $\Gamma$ does not disappear at $T=0$, correspond
to the highest value of $B_{z}$ in the multiplet. There are $k=2$
and $k=3$ tunneling multiplets seen in Fig. \ref{Fig:Gamma-3d-Javier}.
Note that tunneling via low-lying resonances is relatively weak and
it is eclipsed by the thermal activation contribution at higher temperatures.

At stronger transverse field such as $B_{\bot}=3.5$T in Fig. \ref{Fig:Gamma-3d-Bz=00003D3.5},
the barrier is strongly reduced and high-lying tunneling resonances
are broadened away. Here, one can see the ground-state resonance ($B_{z}=0.522$T)
and the first-excited-state resonance ($B_{z}=0.490$T) for $k=1$
multiplet. The ground-state resonance does not disappear at the highest
temperature that has an important implication in the dynamics of fronts
of tunneling. Note the much higher tunneling rate at $T=0$, in comparison
to the previous figure.

A puzzle in the theory of relaxation of molecular magnets is the prefactor
$\Gamma_{0}$ in the Arrhenius relaxation rate, Eq. (\ref{eq:GammaDef}),
being by two orders of magnitude too small. This was already recognized
in the early work \cite{garchu97prb}. Using the standard spin-lattice
relaxation model considering one spin in an infinite elastic matrix,
it is impossible to arrive at $\Gamma_{0}\simeq10^{7}s^{\lyxmathsym{\textminus}1}$
observed in experiments \cite{luisetal97prb,gometal98prb} without
introducing artificially strong spin-phonon interactions \cite{leulos99epl}.
For a strongly diluted molecular magnet, considering a single spin
could be justified, but in the regular case it can not. High density
of magnetic molecules should lead to such collective effects as superradiance
\cite{dic54,chugar02prl,chugar04prl} and phonon bottleneck \cite{abrble70,gar07prb,gar08prb}.
Possibility of superradiance in fast avalanches triggered by a fast
field sweep has been discussed in Ref. \cite{vanetal04prbrc}. References
\cite{heretal05epl,oreker09prb} report microwave emission from MM
that can be interpreted as superradiance. However, it would be difficult
to address these complicated issues while dealing with the quantum
deflagration problem, so that the calculated relaxation rate will
be simply multiplied by 100 to approximately match the experiment,
as was done in Ref. \cite{garsho12prb}.

\subsection{Dipolar field in molecular magnets\label{sub:Dipolar-field}}

Very sharp resonance peaks in the relaxation rate $\Gamma$ seen in
Figs. \ref{Fig:Gamma-3d-Javier} and \ref{Fig:Gamma-3d-Bz=00003D3.5}
require an accurate calculation of the dipolar field in the crystal
that can self-consistently control tunneling by setting individual
molecules on or off resonance. The equations describing this are the
same relaxational equation (\ref{eq:ndotEq}) and thermal equation
(\ref{eq:EdotEq}), as before, only with $\Gamma$ depending on the
total magnetic field
\begin{equation}
B_{\mathrm{tot},z}(\mathbf{r})=B_{z}+B_{z}^{(D)}(\mathbf{r}),\label{eq:Btot_z}
\end{equation}
 where $B_{z}$ is the external bias field and $B_{z}^{(D)}$ is the
self-consistently calculated dipolar field. In the case of cold deflagration,
the thermal equation can be discarded and one has to solve only the
relaxational equation (\ref{eq:n_rate_Eq}). Since the dipolar field
depends on the magnetization everywhere in the crystal, the equations
of quantum deflagration are integro-differential equations. Note that
the transverse component of the dipolar field can be discarded because
its effect is small.

For the purpose of calculating the dipolar field, conventional magnetostatics
(see, e.g., Ref. \cite{lanlif8}) is unsuitable because it provides
an irrelevant magnetic field formally averaged over the microscopic
scale that ignores the lattice structure. The physically relevant
dipolar field is the field created at positions of magnetic molecules
by all other molecules. It is a microscopic quantity that depends
on the lattice structure. To illustrate this point, magnetostatic
field in a uniformly magnetized long sample is $\mathcal{B}^{(D)}=4\pi M$,
where $M$ is the magnetization. However, microscopically calculated
dipolar field in a long uniformly magnetized crystal of Mn$_{12}$Ac
is much smaller, $B_{z}^{(D)}=5.26M$.

It is convenient to express the $z$ component of dipolar field at
site $i$ (i.e., at a particular magnetic molecule) in the form
\begin{equation}
B_{z}^{(D)}=\left(Sg\mu_{B}/v_{0}\right)D_{zz},\label{eq:Bz_via_Dzz}
\end{equation}
where $v_{0}$ is the unit-cell volume. For Mn$_{12}$Ac one has $Sg\mu_{B}/v_{0}=5.0$
mT. The reduced dipolar field $D_{zz}$, created by all other molecular
spins polarized along the $z$ axis is given by
\begin{equation}
D_{i,zz}\equiv\sum_{j}\phi_{ij}\sigma_{jz},\qquad\phi_{ij}=v_{0}\frac{3\left(\mathbf{e}_{z}\cdot\mathbf{n}_{ij}\right)^{2}-1}{r_{ij}^{3}},\qquad\mathbf{n}_{ij}\equiv\frac{\mathbf{r}_{ij}}{r_{ij}},\label{eq:Dzz_Def}
\end{equation}
where $\sigma_{z}\equiv S_{z}/S$. To calculate the sum over the lattice
for the site $i$, one can introduce a small sphere of radius $r_{0}$
around $i$ satisfying $v_{0}^{1/3}\ll r_{0}\ll L$, where $L$ is
the (macrocopic) size of the sample. The field from the spins at sites
$j$ inside this sphere can be calculated by direct summation over
the lattice, whereas the field from the spins outside the sphere can
be obtained by integration. The sum of the two contributions does
not depend of $r_{0}$. If the magnetization in the crystal depends
only on the coordinate $z$ along the symmetry axis of the crystal
that coincides with the magnetic easy axis $z$ (that is the case
for a flat deflagration front), the integral over the volume can be
expressed via the integral over the crystal surfaces. The corresponding
contribution can be interpreted as that of molecular currents flowing
on the surface. The details are given in the Appendix to Ref. \cite{garchu08prb}.

In particular, for a uniformly magnetized ellipsoid the total result
has the form
\begin{equation}
D_{zz}\equiv\sigma_{z}\sum_{j}\phi_{ij}=\bar{D}_{zz}\sigma_{z},\label{eq:Dzz_Ellipsoid}
\end{equation}
independently of $i,$ where
\begin{equation}
\bar{D}_{zz}=\bar{D}_{zz}^{(\mathrm{sph})}+4\pi\nu\left(1/3-n^{(z)}\right)\label{eq:Dzz_Ellipsoid-final}
\end{equation}
and $\nu$ is the number of molecules per unit cell ($\nu=2$ for
Mn$_{12}$Ac having a body-centered tetragonal lattice). Here $\bar{D}_{zz}^{(\mathrm{sph})}$
comes from the summation over a small sphere and the remaining terms
come from the integration. For the demagnetizing factor one has $n^{(z)}=0,$
$1/3,$ and 1 for a cylinder, sphere, and disc, respectively. One
obtains $\bar{D}_{zz}^{(\mathrm{sph})}=0$ for a simple cubic lattice,
$\bar{D}_{zz}^{(\mathrm{sph})}<0$ for a tetragonal lattice with $a=b<c$,
and $\bar{D}_{zz}^{(\mathrm{sph})}>0$ for that with $a=b>c.$ The
latter is the case for Mn$_{12}$Ac having $\bar{D}_{zz}^{(\mathrm{sph})}=2.155$.
For a long cylinder this results in $\bar{D}_{zz}^{(\mathrm{cyl})}=10.53$
or, in real units \cite{garchu08prb,mchughetal09prb},
\begin{equation}
B_{z}^{(D)}=52.6\,\textrm{mT}.\label{eq:BD-value}
\end{equation}

The dipolar energy per magnetic molecule can be written in the form
$E_{0}=-(1/2)\bar{D}_{zz}E_{D}$, where
\begin{equation}
E_{D}\equiv\left(Sg\mu_{B}\right)^{2}/v_{0}\label{eq:ED_Def}
\end{equation}
is the characteristic dipolar energy, $E_{D}/k_{B}=0.0671$ K for
Mn$_{12}$Ac. The role of the DDI in spin tunneling is defined by
the ratio of the typical dipolar bias $W^{(D)}=2Sg\mu_{B}B_{z}^{(D)}=2E_{D}\bar{D}_{zz}^{(\mathrm{cyl})}$
to the width of the overdamped tunneling resonance $\Gamma_{m'}$
in Eq. (\ref{eq:Gamma_resonant}). It is thus convenient to introduce
the parameter
\begin{equation}
\tilde{E}_{D}\equiv2E_{D}/(\hbar\Gamma_{m'})\label{eq:ED_til_Def}
\end{equation}
that is always large. For instance, using the experimental Arrhenius
prefactor $\Gamma_{0}\simeq10^{7}s^{\lyxmathsym{\textminus}1}$ for
$\Gamma_{m'}$, one obtains $\tilde{E}_{D}\simeq10^{3}$.

\begin{figure}[t]
\centering\includegraphics[angle=-90,width=8cm]{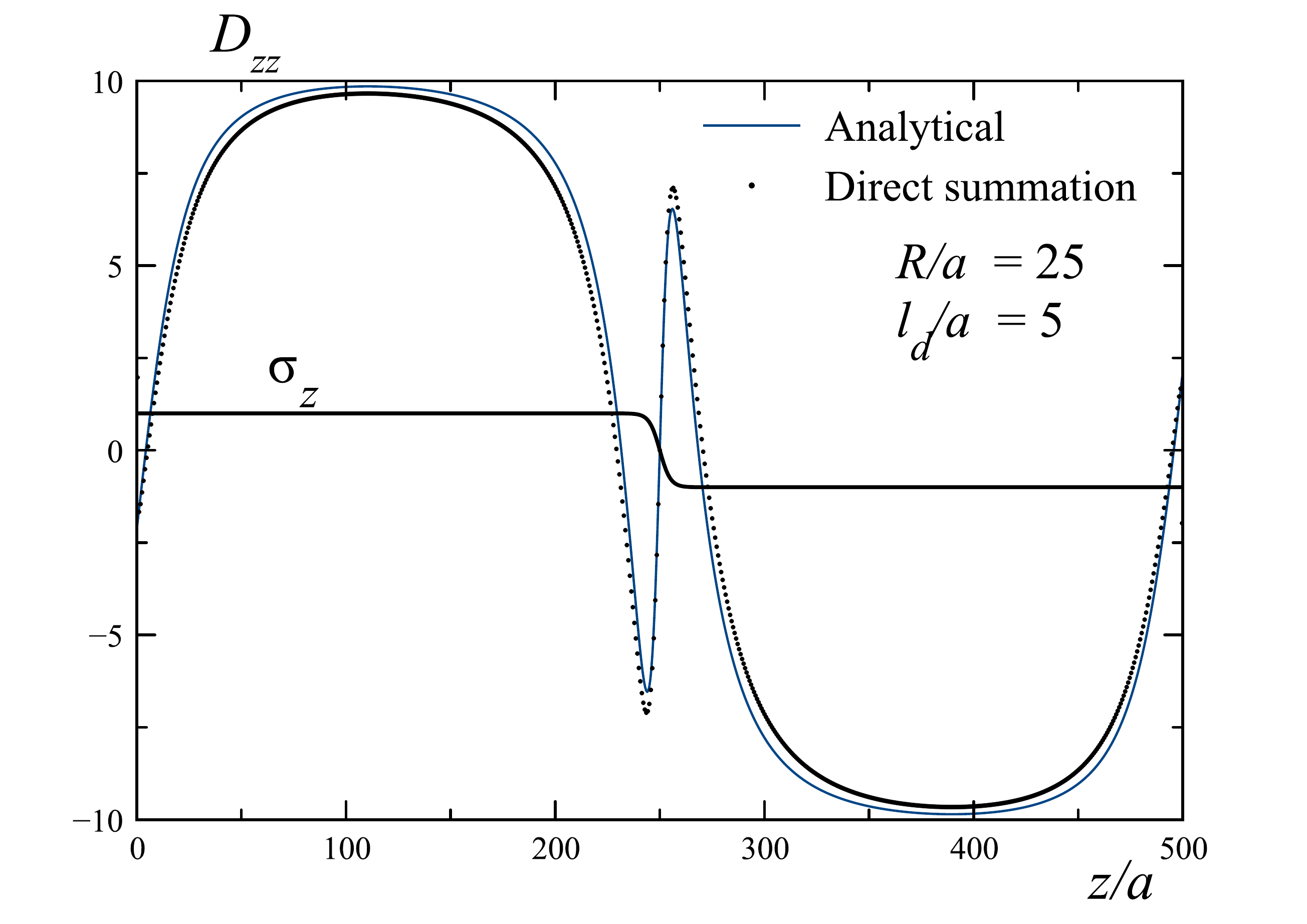}

\caption{Reduced dipolar field in a deflagration front in the slow-burning
limit, created by the magnetization profile $\sigma_{z}(z)=-\tanh\left[(z-z_{0})/l_{d}\right]$.
Analytical result: Eq. (\ref{eq:Dzz_Cylinder}); Points: Direct summation
of dipolar fields over Mn$_{12}$Ac lattice.}

\label{Fig:Dzz_profile}
\end{figure}

For a cylinder of length $L$ and radius $R$ with the symmetry axis
$z$ along the easy axis, magnetized with $\sigma_{z}=$ $\sigma_{z}(z),$
the reduced dipolar field along the symmetry axis has the form \cite{garchu08prb}
\begin{equation}
D_{zz}(z)=\int_{-L/2}^{L/2}dz^{\prime}\frac{2\pi\nu R^{2}\sigma_{z}(z^{\prime})}{\left[\left(z^{\prime}-z\right)^{2}+R^{2}\right]^{3/2}}-k_{D}\sigma_{z}(z),\label{eq:Dzz_Cylinder}
\end{equation}
where $\sigma_{z}=1-2n$ is polarization of pseudospins representing
spins of magnetic molecules ($\sigma_{z}=\pm1$ in the ground and
metastable states, respectively) and
\begin{equation}
k_{D}\equiv8\pi\nu/3-\bar{D}_{zz}^{(\mathrm{sph})}=4\pi\nu-\bar{D}_{zz}^{(\mathrm{cyl})}>0,\label{eq:kDef}
\end{equation}
$k_{D}=14.6$ for Mn$_{12}$Ac. In Eq. (\ref{eq:Dzz_Cylinder}), the
integral term is the contribution of the crystal surfaces, while the
lattice-dependent local term is the contribution obtained by direct
summation over lattice site within the small sphere $r_{0}$ minus
the integral over this sphere that must be subtracted from the integral
over the whole crystal's volume. For other shapes such as elongated
rectangular, one obtains qualitatively similar expressions \cite{gar09prb}.

A striking feature of Eq. (\ref{eq:Dzz_Cylinder}) is that the integral
and local terms have different signs. The integral term changes at
the scale of $R$ while the local term can change faster, that creates
a non-monotonic dependence of $D_{zz}(z)$. In the case of a regular
magnetic deflagration, the spatial magnetization profile in the slow-burning
limit is of the type $\sigma_{z}(z)=-\tanh\left[(z-z_{0})/l_{d}\right]$,
where $l_{d}$ is the width of the deflagration front that satisfies
$l_{d}\ll R$, c.f. Eq. (\ref{eq:nviaxtfront}). The resulting dipolar
field is shown in Fig. \ref{Fig:Dzz_profile}, where the line is the
result of Eq. (\ref{eq:Dzz_Cylinder}) and points represent the dipolar
field along the symmetry axis of a long cylindrical crystal calculated
by direct summation of microscopic dipolar fields over the Mn$_{12}$Ac
lattice. One can see that Eq. (\ref{eq:Dzz_Cylinder}) is pretty accurate,
small discrepancies resulting from $l_{d}$ being not large enough
in comparizon to the lattice spacing $a.$ The central region with
the large positive slope is dominated by the local term of Eq. (\ref{eq:Dzz_Cylinder})
that changes in the direction opposite to that of the magnetization.
For $R\ggg l_{d}$, $D_{zz}$ reaches the values $\pm14.6$ due to
the local term before it begins to slowly change in the opposite direction.
In real units the dipolar field at the local maximum and minimum is
$\pm B_{z}^{(k_{D})}$, where
\begin{equation}
B_{z}^{(k_{D})}=72.9\,\mathrm{mT,}\label{eq:BzkD}
\end{equation}
exceeding the dipolar field of the uniformly magnetized long cylinder
Eq. (\ref{eq:BD-value}). Also one can see from Fig. \ref{eq:Dzz_Cylinder}
that the dipolar field becomes opposite to the magnetization at the
ends of the cylinder, that should lead to an instability of the uniformly-magnetized
state in zero external field.

The $1d$ theory of fronts of tunneling \cite{garchu09prl,gar09prb,garjaa10prbrc,garsho12prb}
is based on the simplifying assumption that the deflagration front
is flat, $\sigma_{z}=\sigma_{z}(z)$, and the dipolar field is given
by Eq. (\ref{eq:Dzz_Cylinder}) everywhere. Since, in fact, the dipolar
field also depends on the distance from the crystal's symmetry axis,
it is likely that such a more complicated structure of $B_{z}$ will
self-consistently affect the front structure, making it non-flat.

There is also a question of stability of a smooth front at a small
scale. Whereas flat and smooth fronts of regular burning are stable,
there is an instability mechanism for a flat front in the presence
of tunneling controlled by dipolar fields that will be explained below.
This is why it is important to develop a full $3d$ theory of fronts
of tunneling.

If the magnetization $\sigma_{z}$ of a MM crystal depends on all
the coordinates $x,y,z$ but this dependence still has a macroscopic
scale, one can again use the method of calculating the dipolar field
that combines summation over a small sphere (where $\sigma_{z}$ does
not change) and integration over the remaining volume of the crystal.
In this case the integral over the volume does not reduce to an integral
over the surface and it has to be done numerically. In the solution
of the deflagration problem, it is convenient to discretize the volume
of the crystal and use the same grid to sample the magnetization variables
and to calculate the dipolar field. A problem with this integral is
that a small sphere of radius $r_{0}$ (around earch point $\mathbf{r\equiv r}_{i}$
where the dipolar field is calculated) has to be excluded from integration
and the contribution of this excluded region is comparable with the
total result because of the singularity of the DDI.

The solution to this problem is, for any point $\mathbf{r}$, to add
and subtract the dipolar field in a uniformly magnetized crystal with
$\sigma_{z}=\sigma_{z}(\mathbf{r})$. The total reduced dipolar field
can be thus represented as
\begin{equation}
D_{zz}(\mathbf{r})=\int d\mathbf{r}'\phi(\mathbf{r}'-\mathbf{r})\left(\sigma_{z}(\mathbf{r}')-\sigma_{z}(\mathbf{r})\right)+\sigma_{z}(\mathbf{r})\left(\bar{\mathcal{D}}_{zz}(\mathbf{r})-k_{D}\right),\label{eq:Dzz-3d-Def}
\end{equation}
where $\phi$ is defined in Eq. (\ref{eq:Dzz_Def}). Because of the
terms subtraction at $\mathbf{r}'\rightarrow\mathbf{r}$, the contribution
of the excluded small sphere in the intergal is negligible and the
integral can be extended to the whole volume of the crystal. Then
the values of the integral for all points of a rectangular grid can
be computed via a summation method based on the fast Fourier transform
(FFT) that takes $\sim N\log(N)$ operations, where $N$ is the number
of grid points. Straightforward calculation of the integral costs
$\sim N{}^{2}$ operations and it has to be avoided.

The remainder of Eq. (\ref{eq:Dzz-3d-Def}) corresponds to a uniformly
magnetized crystal and its structure is similar to Eq. (\ref{eq:Dzz_Cylinder}).
Again, the term with $k_{D}$ is the local contribution, while $\bar{\mathcal{D}}_{zz}(\mathbf{r})$
is the contribution of surface molecular currents, the result of conventional
magnetostatics. For a crystal of a rectangular shape with dimensions
$L_{x}\times L_{y}\times L_{z}$ the result can be obtained as a particular
case of Eq. (88) of Ref. \cite{garsch05prb} and it has the form
\begin{equation}
\bar{\mathcal{D}}_{zz}(\mathbf{r})=\sum_{\eta_{x},\eta_{y},\eta_{z}=\pm1}\arctan\frac{\left(L_{x}+\eta_{x}x\right)^{-1}\left(L_{y}+\eta_{y}y\right)\left(L_{z}+\eta_{z}z\right)}{\sqrt{\left(L_{x}+\eta_{x}x\right)^{2}+\left(L_{y}+\eta_{y}y\right)^{2}+\left(L_{z}+\eta_{z}z\right)^{2}}}+\left(x\Rightarrow y\right),\label{eq:Dzz-3d-Surface}
\end{equation}
in total 16 different $\arctan$ terms.

\subsection{Fronts of tunneling at $T=0$\label{sub:Cold-Deflagration}}

The theory of dipolar-controlled fronts of tunneling at $T=0$ (``cold
deflagration'') \cite{garchu09prl,gar09prb} uses the relaxational
equation (\ref{eq:n_rate_Eq}) with the resonance tunneling rate of
Eq. (\ref{eq:Gamma_resonant}), in which the energy bias $W$ is given
by Eq. (\ref{eq:W-bias-Def}) with $B_{\mathrm{tot},z}$ of Eq. (\ref{eq:Btot_z}).
Within the $1d$ approximation \cite{garchu09prl,gar09prb}, the dipolar
field is given by Eqs. (\ref{eq:Bz_via_Dzz}) and (\ref{eq:Dzz_Cylinder})
for a cylinder. The problem is thus an integro-differential equation.

It is convenient to use the reduced energy bias $\tilde{W}\equiv W/\left(2E_{D}\right)$
that has the form
\begin{equation}
\tilde{W}=\tilde{W}_{\mathrm{ext}}+D_{zz},\qquad\tilde{W}_{\mathrm{ext}}=\frac{(S+m')g\mu_{B}}{2E_{D}}(B_{z}-B_{k}),\label{eq:W_tilde-Def}
\end{equation}
where $m'=S-k$ is close to $S$ for not too strong bias. Propagating
dipolar-controlled fronts of tunneling have been found numerically
\cite{garchu09prl,gar09prb} and analytically \cite{gar09prb} within
the dipolar window near the resonance
\begin{equation}
0\leq\tilde{W}_{\mathrm{ext}}\leq\bar{D}_{zz}^{(\mathrm{cyl})},\label{eq:Dipolar_window-W}
\end{equation}
where $\bar{D}_{zz}^{(\mathrm{cyl})}=10.53$. In real units this yields
the dipolar window
\begin{equation}
B_{k}\leq B_{z}\leq B_{k}+B_{z}^{(D)},\label{eq:Dipolar_window-Bz}
\end{equation}
where $B_{z}^{(D)}$ is given by Eq. (\ref{eq:BD-value}) for Mn$_{12}$Ac.

The solution for the front of tunneling depends on several parameters
such as the transverse size of the crystal $R$ and the resonant value
of the relaxation rate of Eq. (\ref{eq:Gamma_resonant}), $\Gamma_{\mathrm{res}}=\Delta^{2}/(\hbar^{2}\Gamma_{m'})$.
Rewriting the equations in a reduced form \cite{gar09prb}, one immediately
finds that the front speed is of order $\Gamma_{\mathrm{res}}R$.
The only non-trivial parameter is $\tilde{E}_{D}$, Eq. (\ref{eq:ED_til_Def}).
An analytical solution of the problem is possible because of the large
value of $\tilde{E}_{D}$. The front speed is given by \cite{gar09prb}
\begin{equation}
v=v^{*}\Gamma_{\mathrm{res}}R,\qquad v^{*}\simeq\frac{B_{z}-B_{k}}{B_{k}+B_{z}^{(D)}-B_{z}},\label{eq:vDiverging}
\end{equation}
within the dipolar window, independently of $\tilde{E}_{D}$. Above
$B_{k}+B_{z}^{(D)}$ the front speed is zero. The reason for this
is that for the external field above $B_{k}+B_{z}^{(D)}$, the total
field well before the front (where all spins are directed in the metastable
negative direction and produce the dipolar field $-B_{z}^{(D)}$)
is above its resonance value $B_{k}$ (and spin tunneling would even
increase the total field). Thus in this case resonance tunneling transitions
cannot occur. To the contrast, just below $B_{k}+B_{z}^{(D)}$ the
field well before the front is a little bit below the resonance and
increases closer to the front where the magnetization is switching.
In this case, there is a wide region where the system is close to
the resonance, and the front speed becomes very high. Thus as $B_{z}$
crosses the value $B_{k}+B_{z}^{(D)}$ from below, the front speed
diverges and then drops abruptly.

Let us compare the speed of fronts of tunneling $v\simeq\Gamma_{\mathrm{res}}R$
with the speed of regular deflagration, Eq. (\ref{eq:v_via_vtil}).
With a sufficiently strong transverse field applied, one can have
$\Delta/\hbar\sim\Gamma_{m'}$ at the applicability limit of the overdamped
approximation, and then $\Gamma_{\mathrm{res}}\sim\Gamma_{m'}\gg\Gamma_{f}$
because thermal activation goes over high levels of the magnetic molecule
where the distances between the levels and thus the energies of phonons
involved are much smaller than for the low-lying levels, and also
because $\Gamma_{f}$ is exponentially small since $T_{f}\lesssim U$.
Additionally, estimation of $l_{d}$ with $\kappa_{f}=10^{-5}$m$^{2}$/s
and the experimental value $\Gamma_{0}=10^{7}$s$^{-1}$ yield $l_{d}\sim3\times10^{-4}$
mm for $B_{z}$ near the first tunneling resonance and even smaller
for larger bias. As in the experiment the width of the crystal was
much larger than $l_{d}$ (0.3 mm in Ref. \cite{suzetal05prl}, 0.2
mm in Ref. \cite{mchughetal07prb}, and 1 mm in Ref. \cite{heretal05prl}),
one can see that $\Gamma_{\mathrm{res}}R\gg\Gamma_{f}l_{d}$ is quite
possible in a strong transverse field, and then the front of spin
tunneling is much faster than the front of spin burning. A very conservative
estimation with $\Gamma_{\mathrm{res}}\Rightarrow\Gamma_{0}=10^{7}$s$^{-1}$
and $v^{*}\Rightarrow1$ for the crystal 0.2 mm thick yields $v\sim1000$
m/s. As said above, in a strong transverse field one can have $\Gamma_{\mathrm{res}}\gg\Gamma_{0}$,
so that the speed of a spin-tunneling front can easily surpass the
speed of sound that is about 2000 m/s in molecular magnets (see analysis
in Ref. \cite{gar08prbrc}).

A hallmark of the cold deflagration is residual metastable population
behind the front \cite{gar09prb} that can be rewritten as
\begin{equation}
n_{f}=\left(B_{z}-B{}_{k}\right)/B_{z}^{(D)}\label{eq:nf}
\end{equation}
(here $n=n_{i}=1$ before the front). One can see that the change
of $n$ across the front $\Delta n=1-n_{f}$ goes to zero at the right
border of the dipolar window, $B_{z}=B_{k}+B_{z}^{(D)}$. This reconciles
the situation with the general requirement that the rate of change
of the magnetization of the crystal $\dot{M}$, limited by the tunneling
parameter $\Delta$, remains finite. Indeed,
\begin{equation}
\dot{M}\propto(1-n_{f})v=\Gamma_{\mathrm{res}}R\left(B_{z}-B_{k}\right)/B_{z}^{(D)}\label{eq:Mdot}
\end{equation}
reaches only a finite value $\dot{M}\propto\Gamma_{\mathrm{res}}R$
at the right border of the dipolar window before it drops to zero.

\begin{figure}[t]
\centering\includegraphics[angle=-90,width=8cm]{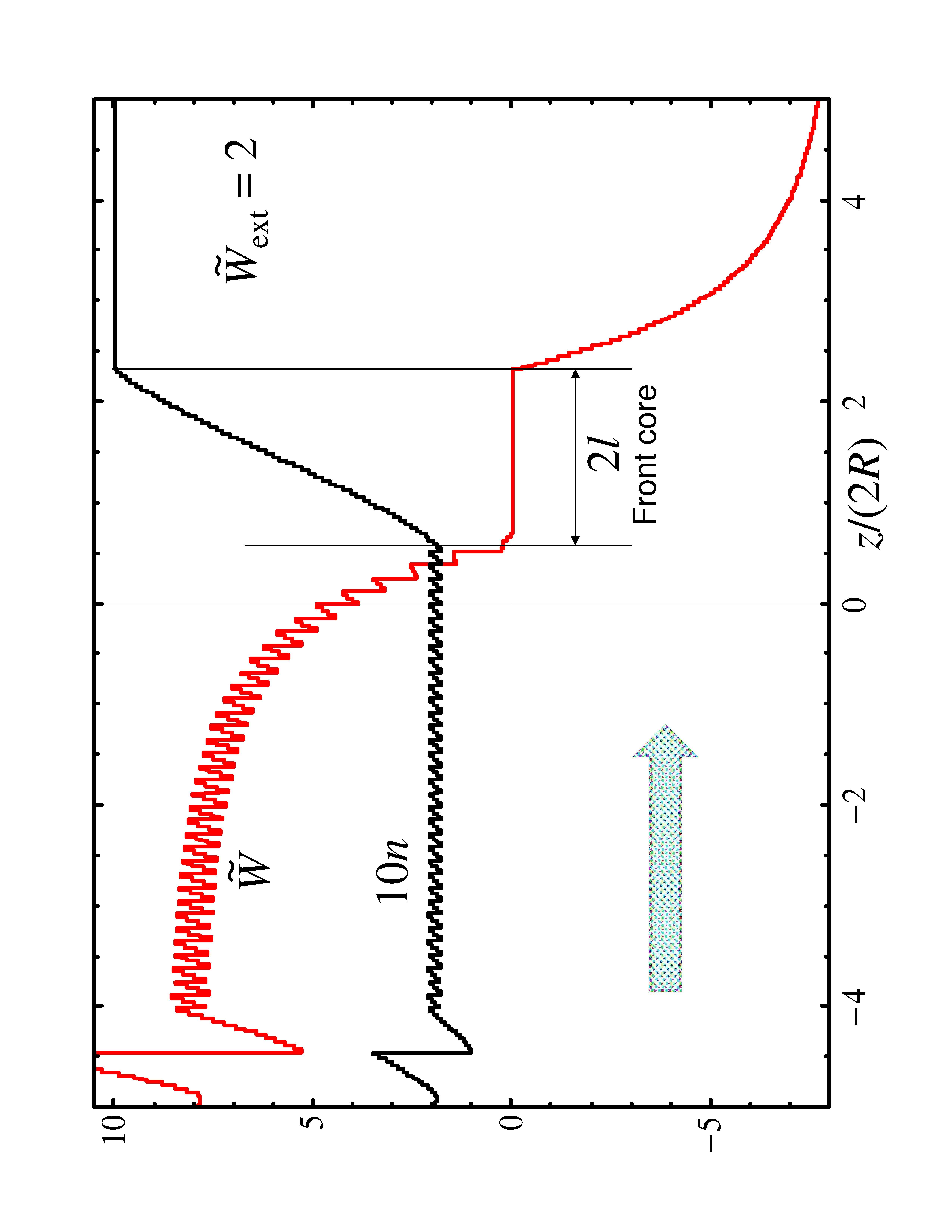}

\caption{Spatial profiles of the metastable population $n$ and the reduced
bias $\tilde{W}$ in the front for $\tilde{W}_{\mathrm{ext}}=2$ and
$\tilde{E}_{D}=20.$ Everywhere in the front the system is near the
resonance, $\tilde{W}\approx0.$ At this value of $\tilde{W}_{\mathrm{ext}}$
the solution begins to lose stability and periodic structures behind
the front begin to emerge. }

\label{Fig:Cold_deflagration_profiles}
\end{figure}

\begin{figure}
\centering\includegraphics[angle=-90,width=8cm]{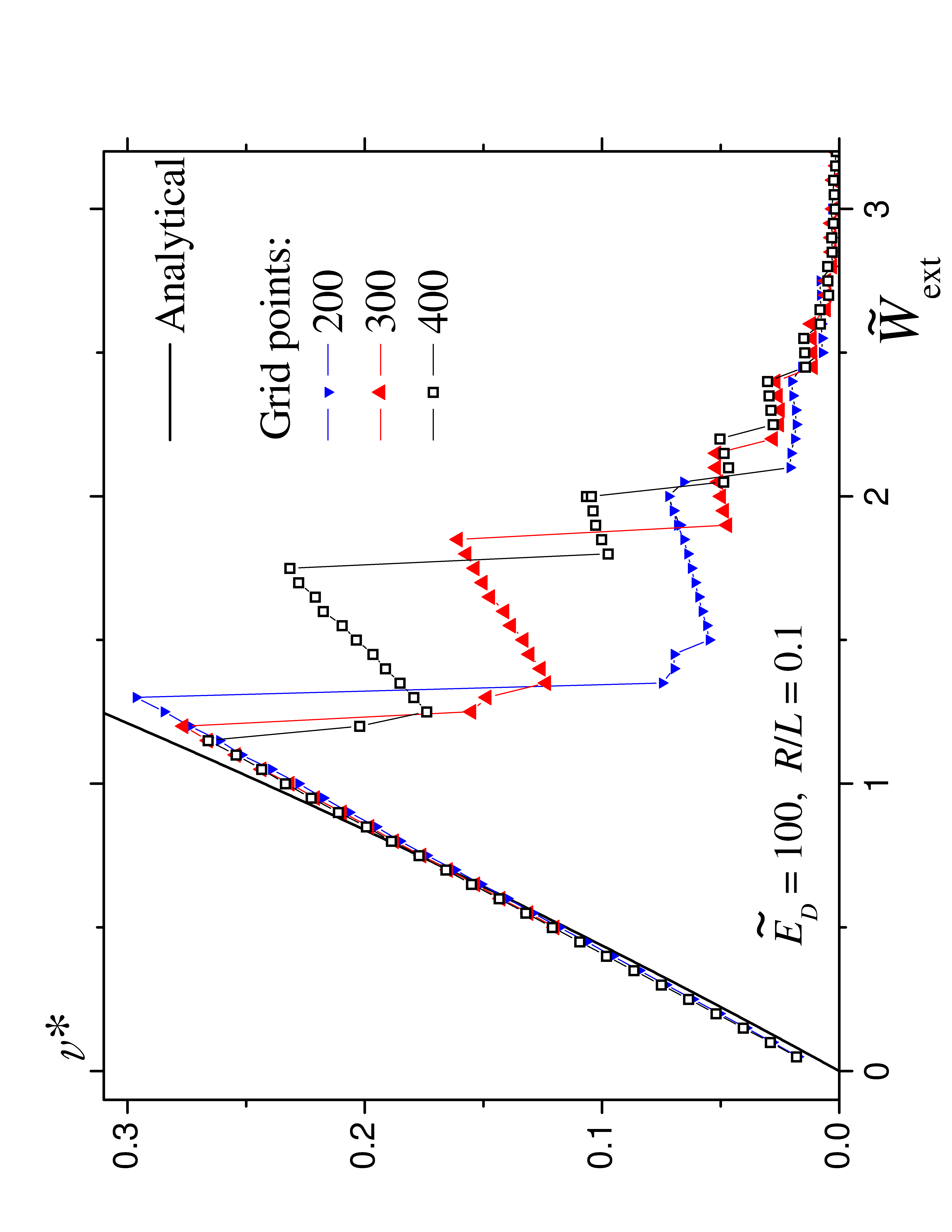}

\caption{Reduced front speed $v^{\ast}$ of Eq. (\ref{eq:vDiverging}) vs the
reduced bias $\tilde{W}_{\mathrm{ext}}$ of Eq. (\ref{eq:W_tilde-Def})
for different number of grid points. For $\tilde{W}_{\mathrm{ext}}\lesssim1$
(the laminar regime) the numerical results are in a good accordance
with Eq. (\ref{eq:vDiverging}) (straight line).}

\label{Fig:vstar_vs_bias_cold}
\end{figure}

To obtain a numerical solution for the cold deflagration, the integro-differential
equation was discretized to make the integral in Eq. (\ref{eq:Dzz_Cylinder})
a sum and the whole problem a set of coupled non-linear first-order
differential equations. The program was written in Wolfram Mathematica.
A typical result for spatial profiles of the metastable population
$n$ and and total energy bias $\tilde{W}$ are shown in Fig. \ref{Fig:Cold_deflagration_profiles}.
In the cold deflagration front, magnetization and dipolar field are
self-consistently adjusting in such a way that inside the front core
of the width $R$ the spins are on resonance and can tunnel. To the
contrary, before and after the front magnetic molecules are off resonance
and tunneling is blocked. One of the reasons why fronts of tunneling
can be so fast is that their width $R$ entering the expression for
the front speed, Eq. (\ref{eq:vDiverging}), is much is much larger
than the width of the deflagration front $l_{d}$, c.f. Eq. (\ref{eq:v_via_vtil}).
The solution shown in Fig. \ref{Fig:vstar_vs_bias_cold} is an example
of the laminar solution for the cold deflagration front that is realized
for a not too strong bias, $\tilde{W}_{\mathrm{ext}}\lesssim$ 1-2
or $B_{z}-B_{k}\lesssim5$-10 mT.

For a stronger bias, the laminar solution becomes unstable. The front
of tunneling is moving with a non-constant speed, leaving spatially-nonuniform
distribution of the unburned metastable population behind. The spatial
dependence of the dipolar field becomes discontinuous and the resonance
condition in the front is not fulfilled (see Fig. 6 of Ref. \cite{gar09prb}).
As a result, the front speed begins to decrease as the instability
develops with the increase of the bias, Fig. \ref{Fig:vstar_vs_bias_cold}.
The instability of the solution is manifesting itself in the dependence
on the discretization, absent in the laminar regime.

The only experimentally feasible method to ignite cold deflagration
is the sweep of the bias field $B_{z}$. When $B_{z}$ is swept in
the positive direction in a negatively magnetized MM crystal, the
resonance condition is first achieved at the ends of the crystal where
the (negative) dipolar field is weaker (see, e.g., the right side
of Fig. \ref{Fig:Dzz_profile}). Spin tunneling at the ends of the
crystal caused by field sweep leads to change of the dipolar field
in this region that brings the system closer to the resonance in a
region of the depth of order $R$, the transverse size of the crystal.
At some moment, a spatial structure close to a stationary front of
tunneling is formed and it begins to propagate into the depth of the
crystal, the field sweep playing no role anymore. This mechanism is
illustrated in Fig. 9 of Ref. \cite{gar09prb}. Numerical calculations
show that front of tunneling is ignited at the ``magic'' value of
the reduced bias $\tilde{W}_{\mathrm{ext}}\simeq5$, weakly dependent
on $\tilde{E}_{D}$. For this value of the bias, the front of tunneling
is non-laminar.

Fronts propagating at other values of the bias, including laminar
fronts, can be ignited by a modified procedure proposed in Ref. \cite{gar09prb}.
First, a global bias is being changed, as before, by a uniform field
sweep until the desired value of $\tilde{W}_{\mathrm{ext}}$ is reached.
After that, front of tunneling can be ignited by a local increase
of the bias near the crystal's end using a small coil producing a
local magnetic field. This method works well in the numerical solution
of the cold deflagration problem. However, such kind of experiment
has not been performed yet.

Cold deflagration can be most likely observed on thinner crystals
having a good thermal contact to the environment, so that the heat
released inside the crystal gets quickly removed and the temperature
does not increase. As said above, the effect only exists within dipolar
windows near tunneling resonances.

It was shown that disorder in resonance fields of individual magnetic
molecules is compensated for by adjustment of the dipolar field in
the front, so that fronts of tunneling survive \cite{garchu09prl}.

\subsection{$1d$ theory of quantum deflagration\label{sub:Quantum-deflagration}}

Here we consider a more general situation in which the temperature
of the crystal is increasing as the result of the decay of the metastable
state, the case when the crystal is thermally insulated. The decay
process is controlled by both the temperature (for any bias) and by
the dipolar field (near tunneling resonances). The theory of the general
quantum-thermal deflagration includes the relaxation equation (\ref{eq:ndotEq})
and the heat conduction (energy diffision) equation (\ref{eq:EdotEq}),
as well as the expression for the dipolar field (\ref{eq:Dzz_Cylinder})
in the $1d$ approximation. The relaxation rate $\Gamma(T,B_{z})$
was calculated for the generic Mn$_{12}$Ac model (\ref{eq:Ham})
in Ref. \cite{garjaa10prbrc} and for the realistic model of Mn$_{12}$Ac
containing the $-AS_{z}^{4}$ term that splits tunneling resonances
in Ref. \cite{garsho12prb}.

Whereas an analytical solution of this problem has not been found,
its qualitattive features can be well understood and the numerical
solution based on discretization is available. In the case of a zero
or weak transverse field, that was the case in all experiments up
to date, spin tunneling is thermally assisted and it only modifies
the main effect of regular deflagration, resulting in tunneling peaks
in the field dependence of the front speed $v(B_{z})$. As in the
case of regular deflagration, ignition can be achieved by raising
the temperature at an end of the crystal.

\begin{figure}[t]
\centering\includegraphics[angle=-90,width=8cm]{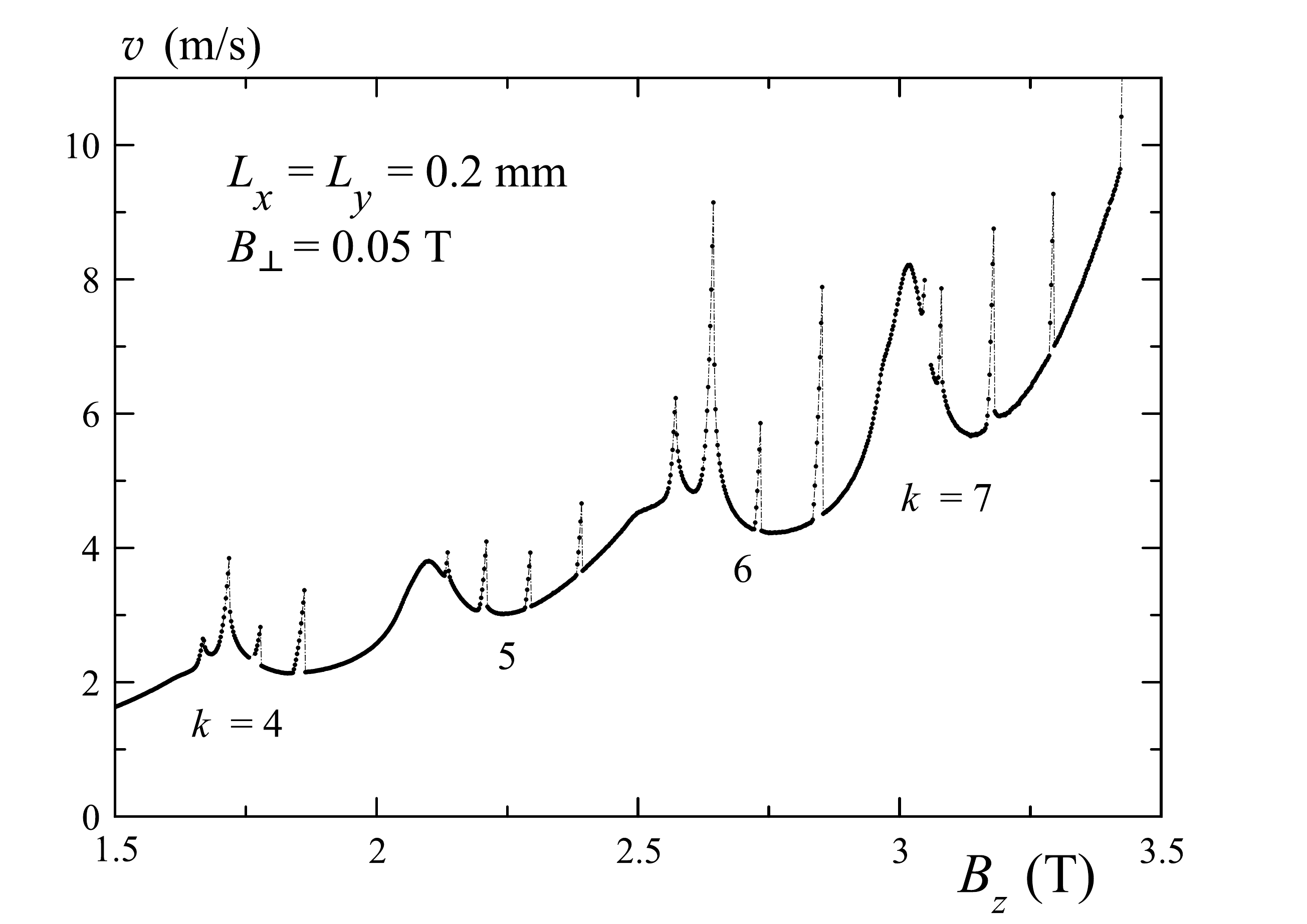}

\caption{Numerically calculated speed of the deflagration front in a long Mn$_{12}$Ac
crystal for a weak transverse field.}

\label{Fig:v_Bz=00003D1.5-3.5T}
\end{figure}

\begin{figure}[t]
\centering\includegraphics[angle=-90,width=8cm]{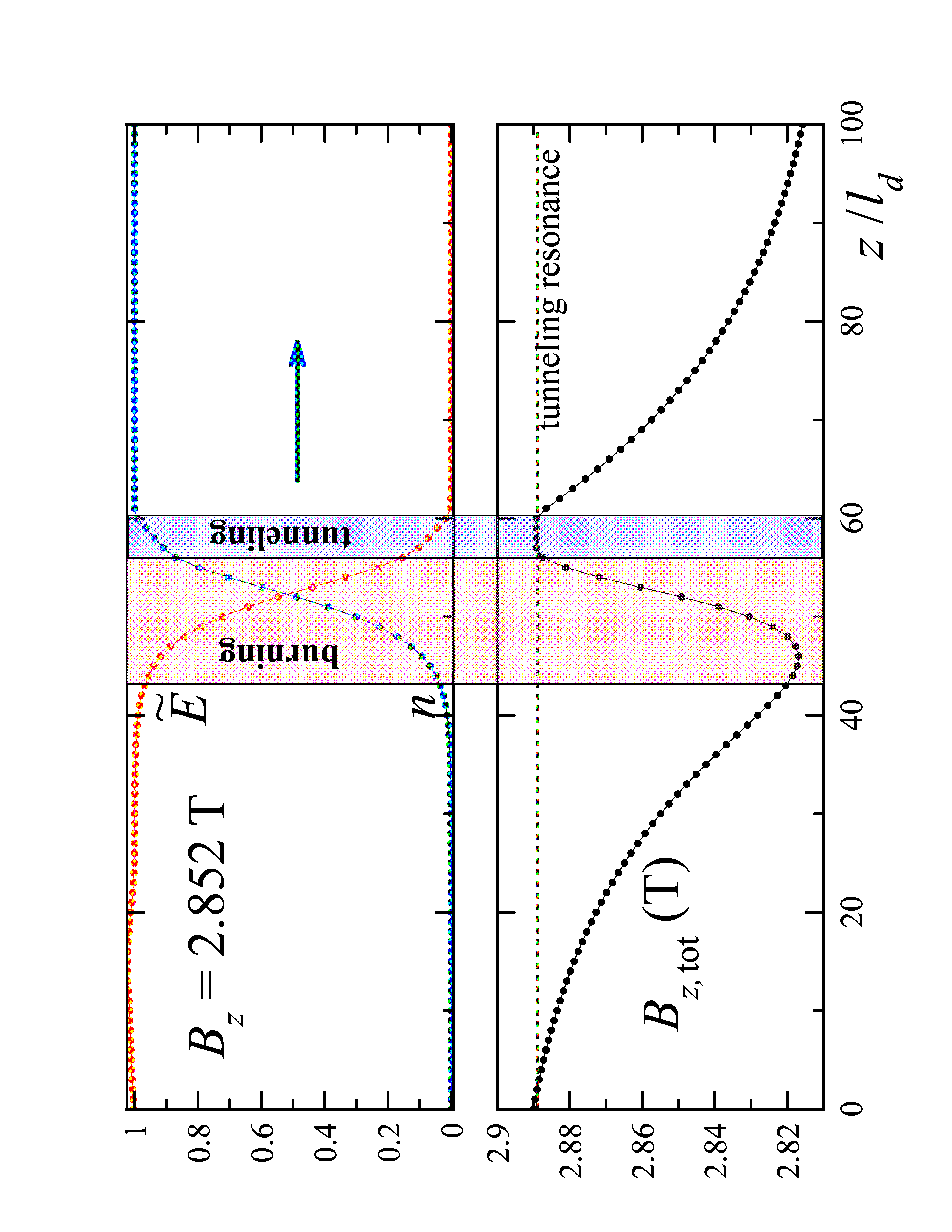}

\caption{Spatial profiles of the deflagration front in a small transverse field,
$B_{\bot}=0.05$ T at the peak of the front speed at $B_{z}=2.852$
T. There is a resonance spin tunneling at the face of the front and
burning in its central and rear parts. In the tunneling region, the
total field $B_{z,\mathrm{tot}}$sticks to its resonance value.}

\label{Fig:Profile-Btr=00003D0.05T_BzExt=00003D2.852T}
\end{figure}

Fig. \ref{Fig:v_Bz=00003D1.5-3.5T} shows the front speed calculated
for the bias and crystal size corresponding to the experiments in
Refs. \cite{mchughetal07prb,hughetal09prb-tuning,hughetal09prb-species}
and using the relaxation rate $\Gamma$ shown in Fig. 3 of Ref. \cite{garsho12prb}.
The tunneling peaks are quite pronounced, at variance with the results
of these experiments. The latter can be due to a large ligand disorder
in Mn$_{12}$Ac that leads to a substantial scatter of the anisotropy
constant $D$ and thus of the positions of the resonances of individual
molecules \cite{parketal01prb,hilletal02prb,parketal02prb}, especially
for the bias as strong as here. Just above 3 T and just below 3.5
T there are regions where the speed is too high to be measured in
this calculation, an effect of ground-state tunneling.

Spatial profiles of the magnetization, energy, and the total bias
field in the deflagration front give an idea of the role played by
spin tunneling. Fig. \ref{Fig:Profile-Btr=00003D0.05T_BzExt=00003D2.852T}
shows the spatial profiles at the asymmetric peak of $v$ at $B_{z}=2.852$
T in Fig. \ref{Fig:v_Bz=00003D1.5-3.5T}. Here the front speed is
high because of tunneling at the face of the front, where in the lower
panel the total bias field is flat at the level of the tunneling resonance
at $B_{z,\mathrm{tot}}=2.889$ T. Magnetization distribution adjusts
so that the dipolar field ensures resonance for a sizable group of
spins that can tunnel. Tunneling of these spins results in energy
release, the temperature and relaxation rate increase, and tunneling
gives way to burning in the central and rear areas of the front.

Formation of the asymmetric maxima of the front speed can be explained
as follows. When $B_{z}$ increases, the peak of $B_{z,\mathrm{tot}}$
that arizes due to the local dipolar field (central part of Fig. \ref{Fig:Dzz_profile})
reaches the resonant value. In thick crystals ($R\gg l_{d}$) this
happens if $B_{z}+B_{z}^{k_{D}}=B_{k}$, where $B_{z}^{k_{D}}$ is
given by Eq. (\ref{eq:BzkD}). This defines the left border of the
dipolar window $B_{z}=B_{k}-B_{z}^{k_{D}}$ (that differs from $B_{z}=B_{k}$
for the cold deflagration). At the left border of the dipolar window,
a strong increase of $v(B_{z})$ begins. The maximum of $B_{z,\mathrm{tot}}$
sticks to the resonance value and becomes flat with progressively
increasing width. Greater width of the resonance region results in
a stronger tunneling and higher front speed. With further increase
of $B_{z}$, the right edge of the tunneling region moves too far
away from the front core into the region where the temperature is
too low. As the tunneling resonance in question is thermally assisted,
it disappears at low temperatures, thus the flat region of $B_{z,\mathrm{tot}}$
cannot spread too far to the right. As a result, the flat configuration
of $B_{z,\mathrm{tot}}$ becomes unstable and suddenly $B_{z,\mathrm{tot}}$
changes to the regular shape of the type shown in Fig. \ref{Fig:Dzz_profile}.

\begin{figure}[t]
\centering\includegraphics[angle=-90,width=8cm]{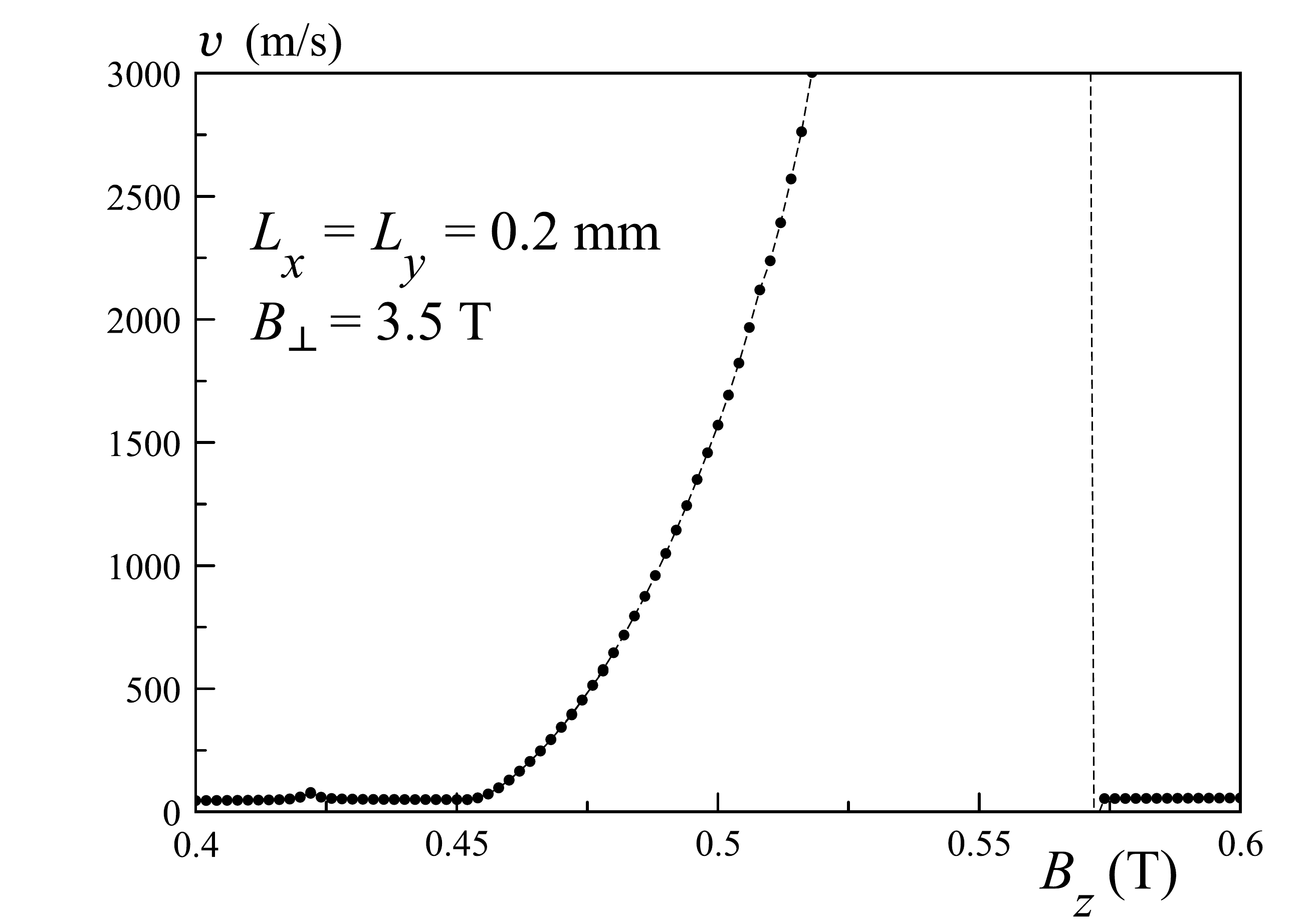}

\caption{Front speed for a strong transverse field ($B_{\bot}=3.5$ T) in the
vicinity of the ground-state tunneling resonance at 0.522 T. The small
peak on the left is due to the first-excited-state tunneling resonance.
Left and right of the dipolar window the front speed is about 50 m/s}

\label{Fig:v_Btr=00003D3.5T_Bz=00003D0.4-0.6T}
\end{figure}

If a strong transverse field is applied, the barrier becomes lower
and it can completely disappear at a ground-state tunneling resonance.
In this case $\Gamma(T,B_{k})$ is practically temperature independent
and this maximum of the relaxation rate does not disappear at the
highest temperatures achieved after burning, $T_{f}$. An example
is the ground-state tunneling maximum at $B_{z}=0.522$T in Fig. \ref{Fig:Gamma-3d-Bz=00003D3.5}.
Although at high temperatures this maximum is hardly visible in the
log scale, it is clearly visible in the normal scale in Fig. 5 of
Ref. \cite{garsho12prb}. In such strong transverse fields, the speed
of the front becomes very high and spin tunneling plays the dominant
role in the front propagation. Figure \ref{Fig:v_Btr=00003D3.5T_Bz=00003D0.4-0.6T}
shows a high front speed within a broad dipolar window
\begin{equation}
B_{k}-B_{z}^{(k_{D})}\leq B_{z}\leq B_{k}+B_{z}^{(D)}\label{eq:Dipolar_window_extended}
\end{equation}
having the width of 125.5 mT. The front speed diverges towards the
right edge of the dipolar window in accordance with Eq. (\ref{eq:vDiverging})
and becomes supersonic. A qualitatively similar behavior was observed
earlier in calculations for the generic model of Mn$_{12}$Ac, see
Fig. 4 of Ref. \cite{garjaa10prbrc}. In contrast to thermally-assisted
tunneling resonances, progressive flattening of $B_{z,\mathrm{tot}}$at
its resonant value is not limited by the temperature before the front
since ground-state tunneling occurs already at zero temperature. Thus
the front speed diverges at the right edge of the dipolar window,
Eq. (\ref{eq:Dipolar_window_extended}), where the width of the tunneling
region becomes very large.

Comparing the present results with the analytical and numerical results
for the cold deflagration, one can see that thermal burning in the
central and rear parts of the front are stabilizing the process, so
that the laminar solution, Eq. (\ref{eq:vDiverging}), holds up to
the right edge of the dipolar window. There is no breakdown of the
laminar regime seen in Fig. \ref{Fig:vstar_vs_bias_cold} at $\tilde{W}_{\mathrm{ext}}\simeq1$.

Another feature of quantum deflagration is complete burning due to
the temperature rise, in contrast to the incomplete burning in the
cold deflagration, Eq. (\ref{eq:nf}). Although the speed of the cold
deflagration front diverges at $B_{z}\rightarrow B_{k}+B_{z}^{(D)}$
(in the laminar regime), the amount of burned metastable population
goes to zero, so that the rate of burning remains finite, Eq. (\ref{eq:Mdot}).
In quantum deflagration burning is complete {[}up to the equilibrium
resudual population $n^{(\mathrm{eq})}$ in Eq. (\ref{eq:neqDef}){]}
while the front speed is diverging, so that the rate of burning is
diverging, too.

Accordingly, the width of the front becomes very large at $B_{z}\rightarrow B_{k}+B_{z}^{(D)}$,
in contrast to the width of the cold-deflagration front that remains
constant. The structure of the front of the quantum-thermal deflagration
near the right border of the dipolar window has a two-tier structure.
First goes a fast front of tunneling that reverts a small fraction
of the magnetization. The latter leads to heat release that ignites
a front of thermal burning that burns all. In the stationary case
the speed of the second part of the front is the same but it takes
time to develop, thus the width of the whole two-tier front is large.
Note that the speed of the quantum deflagration front is not limited
by the speed of sound, contrary to the case of detonation \cite{modbycmar11prl}.

\subsection{$3d$ theory of quantum deflagration\label{sub:QD-3d}}

\begin{figure}[t]
\centering\includegraphics[angle=0,width=8cm]{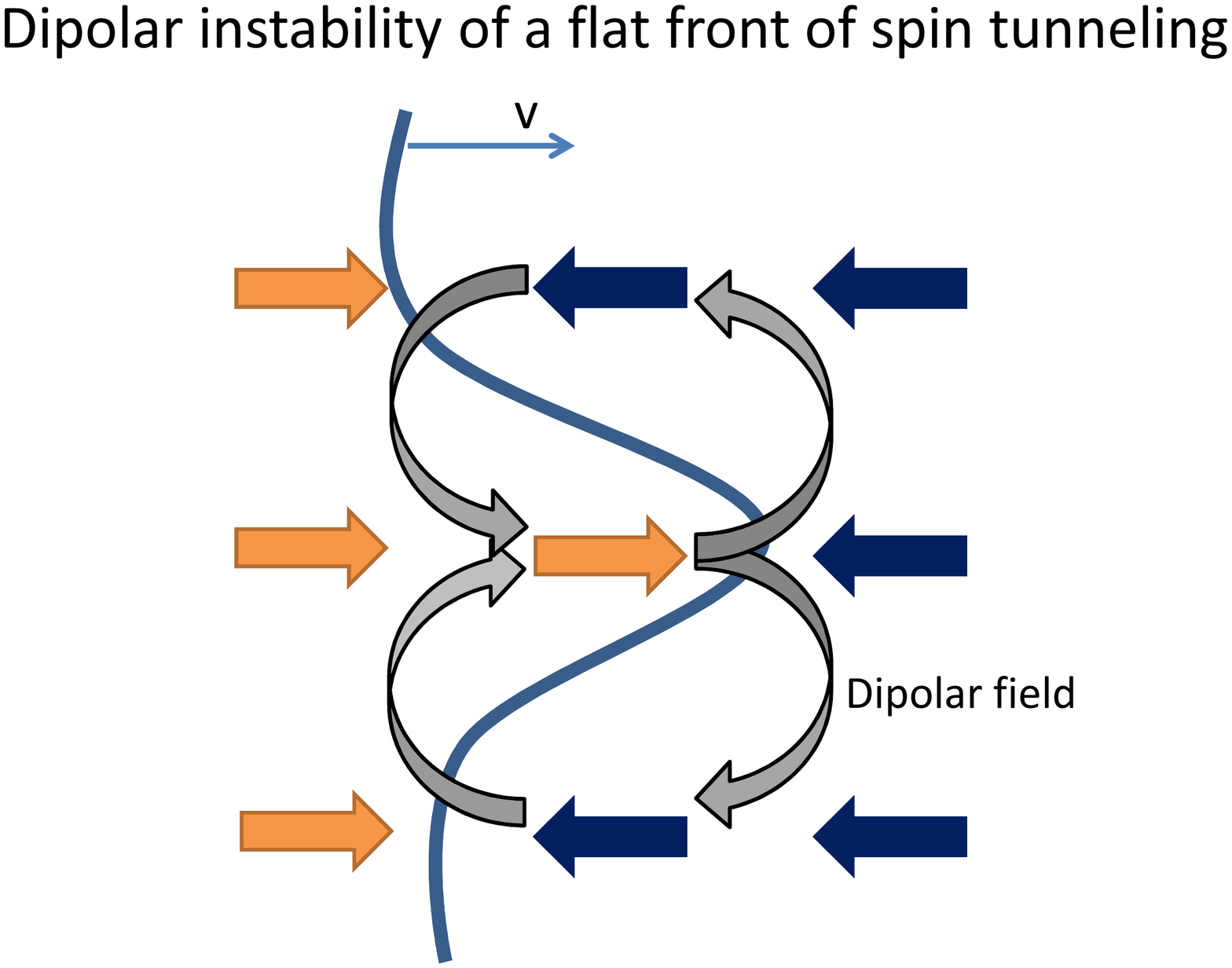}

\caption{Dipolar instability of a flat front of spin tunneling. A leading part
of the front (in the center) produces the dipolar fields on its neighbors
that slow them down.}

\label{Fig:Dipolar_instability}
\end{figure}

As mentioned above, the $1d$ theory of fronts of tunneling assumes
a flat front that is not well justified because the dipolar field
is given by Eq. (\ref{eq:Dzz_Cylinder}) only at the symmetry axis.
Different values of $B_{z,\mathrm{tot}}$ away from the symmetry axis
should self-consistently result in the distribution of the magnetization
that depends on all coordinates $x,y,z$, i.e., in a non-flat front.

On the top of this, there is an instability mechanism for a flat front
at a smaller scale due to DDI. In Fig. \ref{Fig:Profile-Btr=00003D0.05T_BzExt=00003D2.852T}
we have seen that, approaching a front of tunneling from before, $B_{z,\mathrm{tot}}$
increases and reaches the resonance value, then it becomes flat. Now,
if a small fraction of the surface of a front (going from left to
right and changing the magnetization in the positive direction) moves
ahead of its neighbors, it produces a \textit{negative} dipolar field
on the lagging neigboring parts of the front, as any dipole, see Fig.
\ref{Fig:Dipolar_instability}. This brings the neighbors further
from the resonance, so they tunnel later and their lagging increases.
Conversely, lagging portions of the front produce a \textit{positive}
dipolar field on the leading part of the front that helps it to propagate
faster. (The same mechanism leads to instability of flat domain walls
considered in Ref. \cite{garchu08prb}.)

\begin{figure}
\centering\includegraphics[angle=-90,width=8cm]{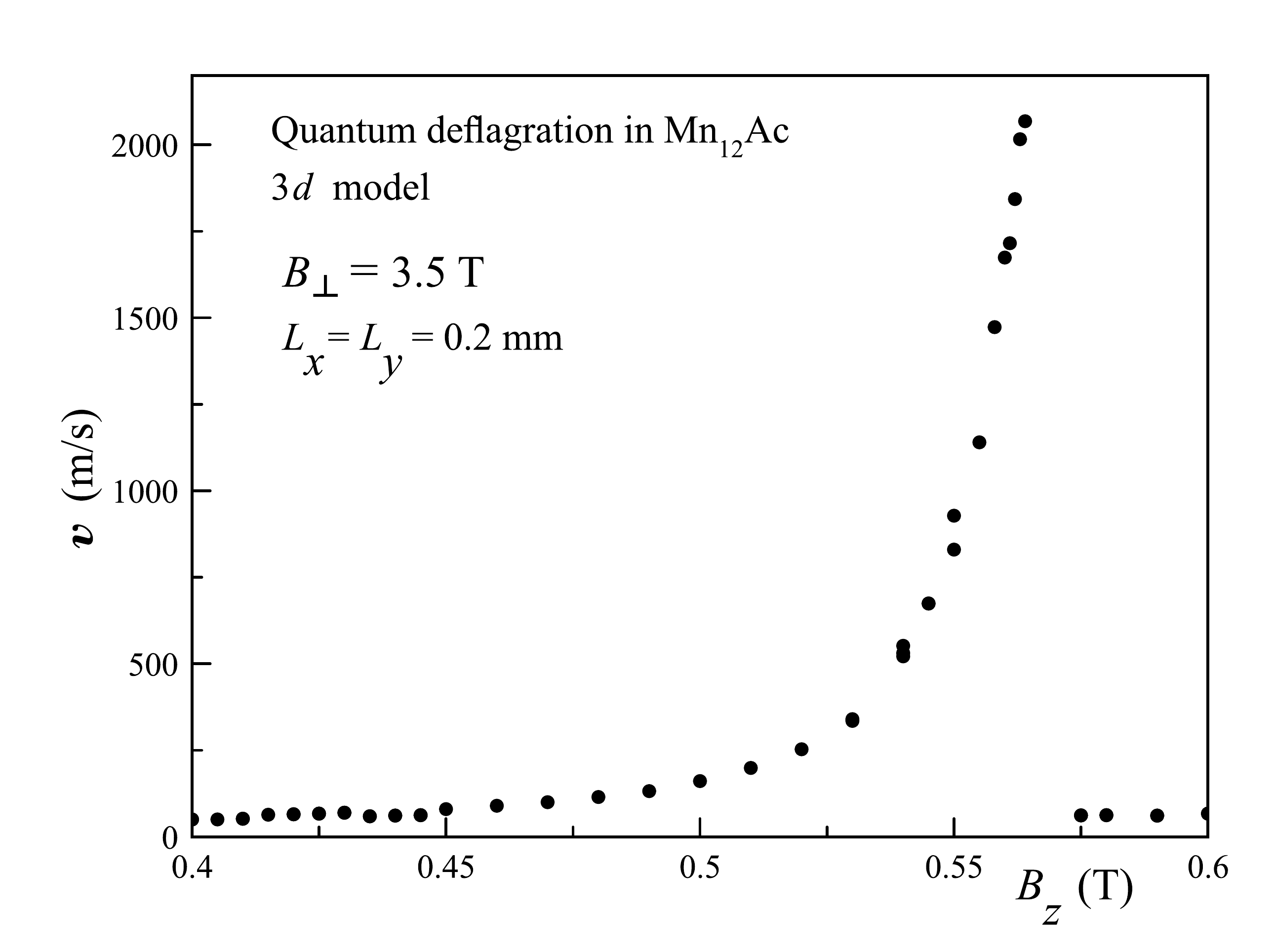}

\caption{Front speed within the $3d$ model for a strong transverse field ($B_{\bot}=3.5$
T) in the vicinity of the ground-state tunneling resonance at $B_{z}=$0.522
T.}

\label{Fig:v_3d-Btr=00003D3.5T_Bz=00003D0.4-0.6T}
\end{figure}

The DDI instability mechanism can potentially destroy any initially
flat front of tunneling, making it microscopically rough. The question
is whether micro-random dipolar fields produced by a micro-random
magnetization in the front are still compatible with resonance tunneling.
It is clear that roughness of the front breaks the concept of the
adjustment of the system to the resonance, so that the speed of the
front should decrease. On the other hand, spins are crossing the resonance,
although at random times, so that still there should be a speed-up
of the deflagration front near tunneling resonances.

\begin{figure}[t]
\centering\includegraphics[angle=-90,width=10cm]{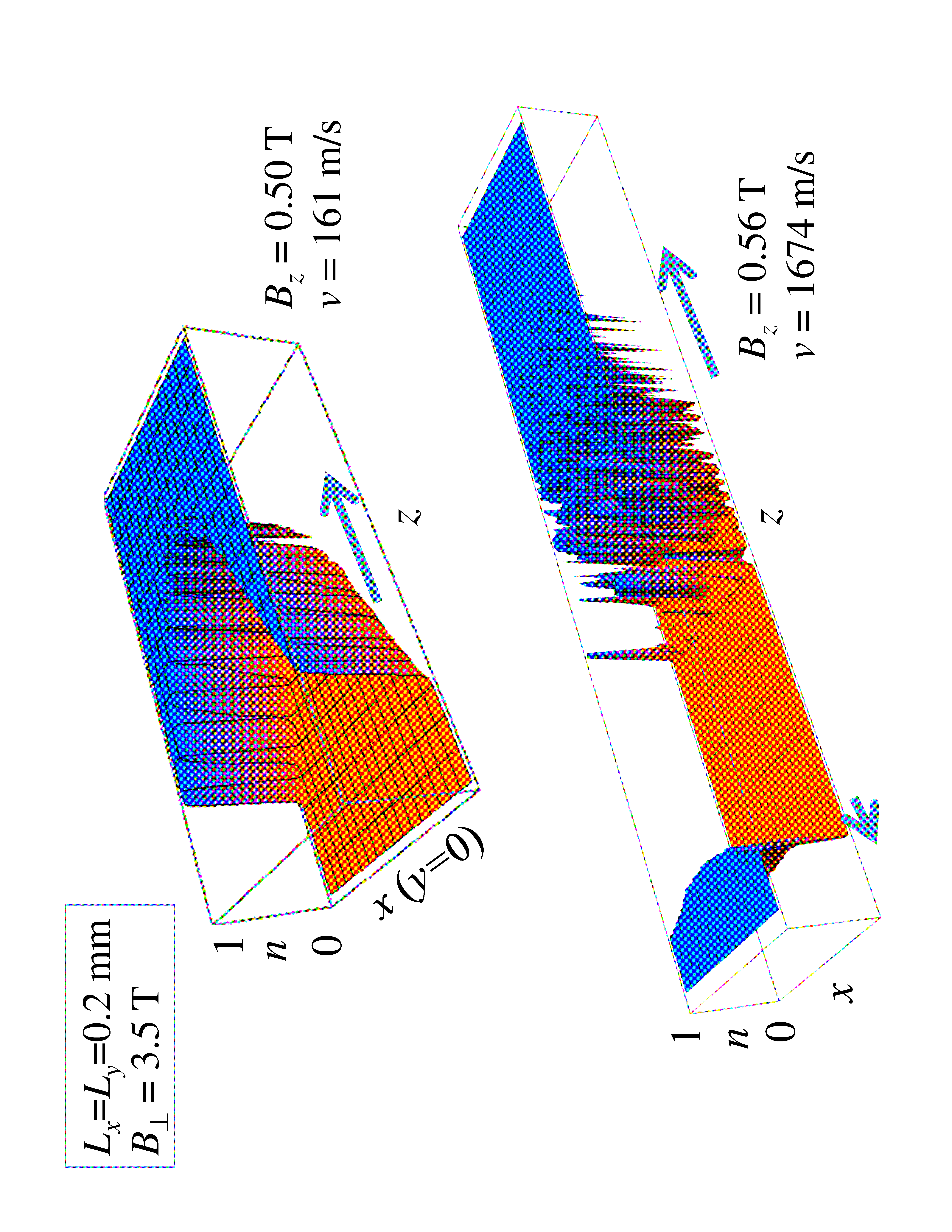}

\caption{Profile of the metastable population $n$ in the $3d$ model of quantum
deflagration for Mn$_{12}$Ac at $B_{\bot}=3.5$T and $B_{z}=0.5$T
(upper) and 0.56T (lower).}

\label{Fig:3d-front_profile_Lx=00003DLy=00003D0.2mm_Bz=00003D0.5_and_0.56T}
\end{figure}

In $3d$ model of quantum deflagration the dipolar field was calculated
using Eq. (\ref{eq:Dzz-3d-Def}) for crystals of box shape with dimensions
$L_{x}=L_{y}\ll L_{z}$ using the relaxation rate $\Gamma$ for $B_{\bot}=3.5$T
shown in Fig. \ref{Fig:Gamma-3d-Bz=00003D3.5}. The crystal was discretized
with about 1 million total grid points in all 3 dimensions. The resulting
system of first-order nonlinear equations was implemented in Wolfram
Mathematica in a vectorized form using a compiled Butcher's 5th-order
Runge-Kutta solver with a fixed step.

As expected, roughness of the front due to the dipolar instability
has been detected within the dipolar window, Eq. (\ref{eq:Dipolar_window_extended}),
where the computed front speed is lower than within $1d$ model, Fig.
\ref{Fig:v_Btr=00003D3.5T_Bz=00003D0.4-0.6T}. Nevertheless, the front
speedup due to spin tunneling is still huge, reaching sonic speeds
in Mn$_{12}$Ac on the right of the dipolar window, see Fig. \ref{Fig:v_3d-Btr=00003D3.5T_Bz=00003D0.4-0.6T}.

Outside the dipolar window, a regular deflagration with a flat front
and front speed $v\simeq50$m/s has been found for this value of the
transverse field. With entering the dipolar window from the left,
the front becomes progressively non-flat with its central part leading.
Front roughness emerges and increases with the bias. Fig. \ref{Fig:3d-front_profile_Lx=00003DLy=00003D0.2mm_Bz=00003D0.5_and_0.56T}
shows the profile of the metastable population $n$ for the crystal
with $L_{x}=L_{y}=0.2$mm, as in experiments of Refs. \cite{mchughetal07prb,hughetal09prb-tuning,hughetal09prb-species},
for $B_{z}=0.5$T and 0.56T. The metastable population $n$ is represented
as a $3d$ plot as a function of $x$ and $z$ with $y=0$ at some
moment of time. The unburned cold portion of the crystal on the right
is shown in blue, while the burned hot part on the left is shown in
red. In the upper part of the figure showing the result for $B_{z}=0.5$T
the front is essentially non-flat and there is some roughness, especially
strong near the symmetry axis. The speed of this front $v=161$m/s
is already much greater than the speed of the regular deflagration,
50m/s.

Numerical results for a larger bias $B_{z}=0.56$T and a longer crystal
are shown in the lower part of Fig. \ref{Fig:3d-front_profile_Lx=00003DLy=00003D0.2mm_Bz=00003D0.5_and_0.56T}.
The front has a nearly sonic speed of $v=1674$m/s and is very rough,
while becoming flat again. The animation of this process looks like
precipitation. Ignition of this front occurs at some distance from
the left end of the crystal where the resonance condition is fulfilled.
From this point, a very fast tunneling front is propagating to the
right while a regular slow burning front is propagating to the left.

\section{Discussion\label{sec:Discussion}}

Regular temperature-driven magnetic deflagration in long crystals
of Mn$_{12}$ has been experimentally observed and is relatively well
understood. The lack of a quantitative accordance between the theory
and experiment can be attributed to still unknown temperature dependence
of the thermal diffusivity $\kappa$, as well as to the absence of
a microscopic theory of relaxation in MM taking into account collective
effects such as phonon/photon superradiance and phonon bottleneck.

Effects of spin tunneling on ignition of deflagration and front speed
near resonance values of the bias field have been experimentally detected
in zero transverse field. However, these effect are due to thermally-assistent
tunneling just below the top of the barrier and they are not strong.

To the contrast, spin tunneling directly out of the metastable ground
state in strong transverse fields can lead to huge effects such as
supersonic quantum deflagration within the dipolar window around tunneling
resonances. Unfortunately, creating an initial state for this process
is practically difficult. In a strong transverse field also \textit{non-resonant}
spin tunneling is rather fast. During the system is being biased to
reach the initial state close to the resonance, it is already relaxing
and a large portion of the metastable population gets lost before
a front of tunneling could start. In addition, non-resonant tunneling
in a biased MM leads to heat release that can result in self-ignition
if the crystal is thermally insulated.

It would be desirable to employ a fast field sweep to bring the MM
into starting position for quantum deflagration in a strong transverse
field without deteriorating its state. To observe non-thermal fronts
of tunneling, thinner crystals with a good thermal contact to the
environment have to be used.

\section{Acknowledgments}

Part of research on magnetic deflagration presented in this Chapter
was conducted jointly with Professor Eugene Chudnovsky. Our students
Reem Jaafar and Saaber Shoyeb participated in obtaining some of the
results. The author is indebted to Ferran Macia, Pradeep Subedi, and
Saül Vélez Centoral for discusions of magnetic deflagration in strong
transverse fields. Oliver Rübenkönig and Daniel Lichtblau have provided
a great support on vectorization and compilation in Wolfram Mathematica.
This work has been supported by research grants from the U.S. National
Science Foundation. This research was supported, in part, under National
Science Foundation Grants CNS-0958379 and CNS-0855217 and the City
University of New York High Performance Computing Center. Eugene Dedits
helped me in using the facilities of the CUNY Computing Center.

\bibliographystyle{spphys}

\end{document}